\documentclass[12pt]{article}
\usepackage[latin1]{inputenc} % devo aggiungere questo pacchetto per avere le lettere
\usepackage[english]{babel}
\usepackage{indentfirst}
\usepackage{fancyhdr}
\usepackage{graphicx}
\usepackage{newlfont}
\usepackage{amssymb} % tutte librerie matematiche
\usepackage{amsmath}
\usepackage{latexsym}
\usepackage{amsthm}
\usepackage{eucal}
\usepackage{eufrak}
\usepackage{bbold}
 \usepackage{tocbibind}
 \usepackage{newlfont}
 \usepackage{enumerate}
 \usepackage{listings}
 \usepackage{tikz}
 \usepackage{bbm}
\usepackage{color}
\usepackage[toc,page]{appendix}
\usepackage{hyperref}

\usepackage[square,sort&compress,numbers]{natbib}
\usepackage{color} %red, green, blue, yellow, cyan, magenta, black, white

\numberwithin{equation}{section}

\definecolor{mygreen}{RGB}{28,172,0} % color values Red, Green, Blue
\definecolor{mylilas}{RGB}{170,55,241}

\theoremstyle{plain}                    %stile corsivo
\newtheorem{theorem}{\bf Theorem}[section]
\newtheorem{lemma}[theorem]{\bf Lemma}
\newtheorem{corollary}[theorem]{\bf Corollary}

\newtheorem{remark}[theorem]{\bf Remark}
\newtheorem{proposition}[theorem]{\bf Proposition}

\newtheorem{assumption}[theorem]{\bf Assumption}

\newcommand*\samethanks[1][\value{footnote}]{\footnotemark[#1]}

\newcommand{\RN}[1]{%
  \textup{\uppercase\expandafter{\romannumeral#1}}%
}

 \title {Liquidity induced asset bubbles via flows of ELMMs}

\author{Francesca Biagini\thanks{Workgroup Financial and Insurance Mathematics, Department of Mathematics, Ludwig-Maximilians Universit{\"a}t, Theresienstraße 39, 80333 Munich, Germany. Emails: biagini@math.lmu.de, meyer-brandis@math.lmu.de} \and Andrea Mazzon\thanks{Gran Sasso Science Institute, viale Francesco Crispi 7, 67100 L'Aquila, Italy. Email:
andrea.mazzon@gssi.infn.it} \and Thilo Meyer-Brandis\samethanks[1]}

\begin{document} 

\maketitle

 \begin{abstract}
We consider a constructive model for asset price bubbles, where the market price $W$ is endogenously determined by the trading activity on the market and the fundamental price $W^F$ is exogenously given, as in \cite{JarrowProtter2012}. To justify $W^F$ from a fundamental point of view, we embed this constructive approach in the martingale theory of bubbles, see  \cite{JarrowProtter2010} and \cite{Biagini}, by showing the existence of a flow of equivalent martingale measures for $W$, under which $W^F$ 
 equals the expectation of the discounted future cash flow. \\
As an application, we study bubble formation and evolution in a financial network. 
\end{abstract}


\lstset{language=Matlab,%
    %basicstyle=\color{red},
    breaklines=true,%
    morekeywords={matlab2tikz},
    keywordstyle=\color{blue},%
    morekeywords=[2]{1}, keywordstyle=[2]{\color{black}},
    identifierstyle=\color{black},%
    stringstyle=\color{mylilas},
    commentstyle=\color{mygreen},%
    showstringspaces=false,%without this there will be a symbol in the places where there is a space
    numbers=left,%
    numberstyle={\tiny \color{black}},% size of the numbers
    numbersep=9pt, % this defines how far the numbers are from the text
    emph=[1]{for,end,break},emphstyle=[1]\color{red}, %some words to emphasise
    %emph=[2]{word1,word2}, emphstyle=[2]{style},    
}

\section{Introduction}

The formation of asset price bubbles has been thoroughly investigated from an economical point of view in many contributions, see Tirole \cite{Tirole}, Allen and Gale \cite{Allen}, Choi and Douady \cite{Douady2009}, \cite{Douady2011}, Harrison and Kreps \cite{HarrisonKreps}, Kaizoji \cite{Kaizoji2000}, Earl et al. \cite{Earl2007}, DeLong, Shleifer, Summers and Waldmann \cite{DeLong}, Scheinkman and Xiong \cite{Scheinkman}, \cite{Scheinkman2013}, Xiong \cite{Xiong}, Abreu and Brunnermeier \cite{Abreu}, F{\"o}llmer, Horst, and Kirman \cite{Follmer}, Miller \cite{Miller}, Zhuk \cite{Zhuk}. \\
Different causes have been indicated as triggering factors for bubble birth, such as heterogenous beliefs between interacting agents (as in \cite{Follmer}, \cite{HarrisonKreps}, \cite{Scheinkman}, \cite{Scheinkman2013}, \cite{Xiong}, \cite{Zhuk}), a breakdown of the dynamic stability of the financial system (\cite{Douady2009}, \cite{Douady2011}), the diffusion of new investment decision rules from a few expert investors to larger population of amateurs (see \cite{Earl2007}), the tendency of traders to choose the same behavior as the other traders' behavior as thoroughly as possible (see \cite{Kaizoji2000}), the presence of short-selling constraints (see \cite{Miller}). \\
From the mathematical point of view, one of the main approaches is given by the martingale theory of bubbles as introduced by Cox and Hobson \cite{CoxHobson} and Loewenstein and Willard \cite{LoewensteinWillard} and mainly developed by Jarrow, Protter et al. \cite{JarrowProtter2009}, \cite{JarrowProtter2011},  \cite{JarrowProtter2007},   \cite{JarrowProtter2010}, \cite{JarrowKchiaProtter}. See Protter \cite{Protter2013} for an overview. 
% Biagini et al. \cite{BiaginiNedelcu}, \cite{Biagini},   Jarrow and Madan \cite{JarrowMadan}, ,  Kardaras et al. \cite{KardarasKreherNikeghbali} a flourishing stream regards the so called martingale theory of bubble formation. 
In this setting a $Q$-bubble is defined as the difference between the market price of a given financial asset and its fundamental value, given by the expectation of the future cash flows under an equivalent local martingale measure $Q$. \\
% This bubble is called $Q$-bubble, and it is therefore defined as
%\begin{equation}\label{Qbubble}
%\beta_t^{Q}=W_t-\mathbb{E}_Q[W_{\tau}|\cal{F}_t], \quad 0\le t \le \tau,  
%\end{equation}
%where $(\cal{F}_t)_{t \in [0,\tau]}$ is the filtration of the probability space taken into consideration and $\tau$ is the maturity, or the liquidation time, of the asset. \\
Defined in this way, the bubble is a non-negative local martingale under $Q$, and it is strictly positive if and only if the market wealth $W$ is a strict $Q$-local martingale (for a complete analysis, see for example \cite{Biagini}, \cite{CoxHobson}, \cite{JarrowProtter2007},   \cite{JarrowProtter2010}, \cite{LoewensteinWillard},   \cite{Protter2013}). \\
In a complete market (see \cite{JarrowProtter2007}), where only one equivalent local martingale measure (ELMM) exists, only two possibilities are given: either no bubble appears at all, or a bubble is already present at the beginning.  This is a strong modeling withdraw, therefore in \cite{JarrowProtter2010} and \cite{Biagini} incomplete markets have been taken into consideration: the birth and the evolution of a bubble are then determined by a flow of different ELMMs that gives rise to a corresponding shifting perception of the fundamental value of the asset. 
%In this approach, the bubble is originated by a change in the perception of the fundamental value in the market, and in particular it takes place when due to a change in the market view there is a switch from a martingale measure  to a measure . 
In \cite{JarrowProtter2010} the underlying pricing measures may change only at certain stopping times, in \cite{Biagini} a continuous flow in the space of martingale measures is considered.\\
On the other hand, an alternative model is given by Jarrow, Protter and Roch in \cite{JarrowProtter2012}, where the fundamental value is exogenously given, whereas the market value is endogenously determined  by the trading activity of investors, and studied through the analysis of the liquidity supply curve. For another constructive model, see also \cite{BiaginiNedelcu}. % where a bubble can take place in the valuation of defaultable claims, whose safety is overestimated by investors under certain circumstances.\\
\\In this setting a  bubble is still defined as the difference between the market value $W$ and the fundamental value $W^F$, however it does not always coincide with the $Q$-bubble under a given equivalent martingale measure $Q$. %This is the case if and only if the fundamental wealth $W^F$ is a martingale under $Q$.\\
%However, in \cite{JarrowProtter2012} only one equivalent martingale measure $Q$ is considered, and the model does not fit the $Q$-bubble approach, since as reported in \cite{JarrowProtter2012} the bubble defined in such a way equals the $Q$-bubble if and only if the fundamental wealth $W^F$ is a martingale under $Q$. \\  
\\ A natural question is then if it is possible to embed a constructive model, where the fundamental price is exogenous and the market price endogenous, in the martingale theory of bubbles, by determining a suitable flow of ELMMs for $W$ under which $W^F$ is justified from a fundamental point of view.\\
More precisely, given a liquidation time $T$ for the financial asset, we look for a flow $(Q^t)_{t \in [0,T)}$ of ELMMs for the market wealth $W$ such that 
%\begin{equation}\label{requirement}
%W_t^F=\mathbb{E}_{Q^t}[W_T^F|\cal{F}_t], \quad t \in [0,T],
%\end{equation}
the fundamental value of the asset is given as the expectation of the future cash flow as in equation (\ref{eqzag}). Note however that we do not obtain that $W^F$ is also a (local) martingale under each measure of the flow, as thoroughly discussed in Remark \ref{remzag}. \\
%In this way, the bubble can be written as 
%$$\beta_t=W_t-W_t^F=W_t-\mathbb{E}_{Q^t}[W_T^F|\cal{F}_t], \quad t \in [0,T].$$
%We remark that (\ref{requirement}) does not imply that $W^F$ is a martingale under every ELMM of the flow, since $Q^t$ represents the pricing measure chosen by the market at time $t$ and eq. (\ref{requirement}) holds $t$-wise.\\
Our main result is then that we can explicitly determine the form of such a flow of ELMMS in a liquidity driven model under very general assumptions, see Theorem \ref{maintheo}. This require a consistent technical effort, mostly devoted to guarantee the martingale property of the chosen flows of (eventual) probability densities. In this way we are able to directly connect the impact of the underlying macro-economic factors to the shift of the resulting pricing measure, which may change over time. \\ %Since eq. (\ref{requirement}) is satisfied, the bubble coincides with the $Q$-bubble for every $t \in [0,T)$. \\
%Therefore, being able to find a flow of ELMMs satisfying (\ref{requirement}), we can  
As an application of our method, we consider the evolution of a bubble in a financial network and compute the generating flow of ELMMs. However, this example is also of independent interest, as it studies how the interaction of market participants in a financial network can affect asset price formation and the consequent birth of a bubble. Different studies show how contagion between investors and herding behavior may play an essential role when a bubble grows up: euphoria and exuberance can propagate among market participants, due to exchanges of ideas (see Lux \cite{Lux}) or to the fact that investors may be attracted by the short period earnings of acquaintances investing in the bubbly asset, as observed by Bayer et al. in \cite{Bayer2014}, where analyzing data from the housing bubble in L. A. in the 2000s the authors notice a strong contagion between neighbors. \\
Several contributions in the last years has been focusing on how some properties of the network, like mean degree or degree heterogeneity, can influence the contagion of failures and losses between banks during a financial crisis (see for example Acemoglu et al. \cite{Acemoglu},  Allen and Gale \cite{AllenGale2000}, Amini et al. \cite{Amini2013}, Cont et al. \cite{Cont2013}, Gai and Kapadia \cite{GaiKapadia}, Newman et al. \cite{NewmanStrogatzWatts}, Watts \cite{Watts}, Watts and Strogatz \cite{WattsStrogatz}). Some investigation has been proposed about how bubbles are generated at the microeconomic level by the interaction of market participants (see among others Lux \cite{Lux}, Scheinkman \cite{Scheinkman},  Scheinkman and Xiong  \cite{Scheinkman2013},  Tirole \cite{Tirole}, Zhuk \cite{Zhuk}). However, only a few studies have been devoted to understand how the structure of a given financial network can influence the spread of contagion between investors that generates a bubble. 
In \cite{Lux}, for example, the author models the bubble as caused by a self-organizing process of infection between traders, expressed by a system of PDEs, leading to equilibrium prices that deviate from the fundamental value. However they consider a world in which everybody is connected with everybody, so that the network structure does not enter into play.  \\
In our special case we focus on a model for the aggregate trading volume of $X$ in dependence by some characteristics of the underlying networks of investors, such as the degree distribution.
In particular we use some modeling approach deriving from the literature on infectious processes in a population by following the so called SIS model (see Pastor-Satorras and Vespignani  \cite{PastorVespignani2001} and \cite{SatorrasVespignani2001}).
We provide numerical simulations to investigate how different networks generate different contagion mechanisms and then to bubbles with different evolutions. In particular, it turns out that in more heterogenous networks (i.e. networks with a more right skewed degree distribution) contagion spread faster at the beginning so that the bubble builds up faster and bursts sooner: the nodes with high degree, which in average get infected faster, contribute with an higher weight in the more right skewed distributions.\\
%In our analysis we look at the process $X$ representing the aggregate trading volume of investors, and express it through an SDEs taking inspiration from the literature on infection processes in a population, and in particular from the so called SIS model (see Pastor-Satorras and Vespignani  \cite{PastorVespignani2001} and \cite{SatorrasVespignani2001}). . \\
The paper is therefore organized as follows: in Section 2 we describe the setting of the liquidity model, define the fundamental value of the asset and specify how the trading activity of investors influences the market price of the asset.  In Section 3 we determine a possible flow $(Q^t)_{t \in [0,T)}$ of ELMMs satisfying (\ref{eqzag}) and show that the density process $(Z_{t,s})_{s \in [0,T)}$ with $Z_{t,s}=\frac{dQ^t}{dP}|_{\cal{F}_s}$ is a true martingale wrt $s$.  In Section 4 we give an example showing how contagion between investors can develop the bubble in a network and compute the generating flow of ELMMs. 
%the authors model the bubble as  In particular, due to liquidity effects (see \cite{JarrowProtter2012} for an exhaustive explanation) $W$ diverges from $W^F$ and a bubble takes place when the aggregate trading volume process $X$ is different from zero, providing the illiquidity $M$ is not zero.\\

\section{The Setting}\label{def}

Let $(\Omega, \cal{F}, P)$ be a probability space and $T>0$ a random time on it, representing the maturity or liquidation time of the underlying risky asset as  in the setting of \cite{JarrowProtter2010}. We assume that $(\Omega, \cal{F}, P)$ is endowed with a filtration $\mathbb{F}=(\cal{F}_t)_{t \in [0,T]}$ satisfying the usual assumptions of completeness and right continuity. \\  %$\mathbb{F}=$ 
On $(\Omega,  \cal{F}, \mathbb{F}, P)$ we have ($B^1,B^2,B^3,B^4,N$), where $B^i=(B^i_t)_{t \in [0,T]}$, $i=1,2,3,4$ are standard $\mathbb{F}$-Brownian motions and $N_t=\mathbb{1}_{\{\tau \le t\}}$  is a jump process with $\tau$ totally inaccessible stopping time with intensity process $\pi=(\pi_t)_{t \in [0,T]}$. We assume that ($B^1,B^2,B^3,B^4,N$) are independent processes. \\
%We have
%\begin{equation}\label{pi}
%\tau=\inf\{s \ge 0: \int_0^s\pi_udu>E\},
%\end{equation}
%where $E$ is an exponential random variable.\\
%Following Jarrow et al. \cite{JarrowProtter2012} we model the bubble as
%$$\beta_t=W_t-W_t^F, \quad 0 \le t \le T,$$ where $T<\infty$ is the liquidation time, $W$ is the market wealth defined by
%$$W_t=D_t+S_t\mathbb{1}_{\{t<T\}}+F_T \mathbb{1}_{\{t=T\}}, \quad 0 \le t \le T$$
%with $D$ dividend process and $F_T$ liquidation value, and $W^F$ is the fundamental wealth process 
%$$W_t^F=D_t+F_t, \quad 0 \le t \le T.$$
Following \cite{JarrowProtter2012} we consider a financial asset whose fundamental wealth $W^F=(W_t^F)_{t \in [0,T]}$ (associated to the cumulative dividend process $(D_t)_{t \in [0,T]}$ and to the liquidation value $F$ of the asset at time $T$) is given by 
\begin{equation}\label{WF}
 dW_t^F= W_t^F (a dt + b dB_t^1), \quad 0 \le t \le T,
 \end{equation}
with $W_0^F>0$, $a>0$ and $b>0$.\\
We interpret $\tau$ as the time of birth of a bubble for this financial asset.
The bubble follows the dynamics
\begin{equation}\label{bubblenew}
d\beta_t=M_t\Lambda_t(-k\beta_tdt+2dX_t+2xW_t^FdN_t), \quad 0 \le t < T,
\end{equation}
where  $X$ is the aggregate trading volume (buy market orders minus sell market orders), $x$ is the aggregate trading volume at $\tau$ and $M=(M_t)_{t \in [0, T]}$, $\Lambda=(\Lambda_t)_{t \in [0, T]}$ are respectively a measure of illiquidity and the so called \textit{resiliency} (for an economical motivation of this setting we refer to \cite{JarrowProtter2012}). We put $\beta_{\tau}=2x\Lambda_{\tau}M_{\tau}W_{\tau}^F$ for a given $x>0$.
 \\%, i.e. the market value jumps at final time $T$, equalizing the fundamental value.\\
We consider that $X$ satisfies the following dynamics
\begin{align}
&X_t=0,  & \text{for }  0 \le t < \tau \notag, \\
 &dX_t=\mu_t dt + \sigma_t dB_t^2,  & \text{ for }  \tau \le t<T, \label{Xgen}
\end{align} 
where $\mu=(\mu_t)_{t \in [0, T]}$ and $\sigma=(\sigma_t)_{t \in [0, T]}$ are progressively measurable processes that a priori can also depend on $X$ itself or on the bubble $\beta$. \\
In \cite{JarrowProtter2012} the aggregate trading volume is modeled as in (\ref{Xgen}) with $\mu \equiv 0$ and $\sigma_t=\alpha\beta_t$. Here we introduce the drift $\mu$ in order to 
see the influence of the network on the size of the bubble, as we specify in Section \ref{secnet}.\\
 Here the fundamental wealth process $W^F$ is exogenously given, while the market wealth process $W=(W_t)_{t \in [0,T]}$ is endogenously determined as 
 $$W_t=W_t^F+\beta_t, \quad 0 \le t < T.$$
At liquidation time $T$ we have $W_T=W_T^F$: the asset is liquidated at time $T$ at the estimated firm's value, i.e. at the fundamental value. In particular we require in the sequel that there exists an equivalent local martingale measure for $W$ only on the open interval $[0,T)$, since around time $T$ the liquidation procedure is not subjected to market equilibrium mechanisms. 

\begin{assumption}\label{remM}

\begin{enumerate}[(i)]
\item $\int_{\tau}^T \mu_s^2ds<\infty$ a.s.
 \item $\int_{\tau}^T \sigma_s^2ds<\infty$ a.s. and $\int_{\tau}^T \frac{1}{\sigma_s^4}ds<\infty$ a.s.
\item $\mu$ and $\sigma$ are such that there exists a unique solution of (\ref{Xgen}) (see for example Theorem 7 in Chapter \rm{V}.3 in \cite{ProtterBook});
%functional Lipschitz in the sense of the Definition in Chapter \rm{V}.3 in \cite{ProtterBook} as functionals of $\beta_t$, so that ). \\ 
 \item $M=(M_t)_{t \in [0, T]}$ is an adapted process that satisfies the dynamics 
\begin{equation}\notag
dM_t = \tilde{\mu}(M_t)dt+\tilde{\sigma}(M_t)dB^3_t, \quad 0 \le t \le T, \label{dM}
\end{equation} 
 where $\tilde{\mu}$ and $\tilde{\sigma}$ are such that there exists a unique solution of (\ref{dM}) according to Theorem 7 in Chapter \rm{V}.3 in \cite{ProtterBook}. Moreover $\int_a^b\tilde{\sigma}^{-4}(x)dx<\infty$ for every $a,b$ such that $0<a<b<\infty$. % and $\int_0^T|\tilde{M_s}|ds<\infty$, $\int_0^T|\tilde{\sigma_s}|^2ds<\infty.$   

%\begin{align}
 %\end{align}  
% \begin{equation}
 %\int_0^t M_s e^{\int_0^s M_u du}ds<\infty \quad \forall t \in [0,T].
 %\end{equation}

% \begin{equation}\label{eqassM}
%dM_t=A_t(\bar{k}-M_t)dt+C_tM_tdB^3_t, \quad 0 \le t \le T
%\end{equation}
%with $\bar{k}>0$, $M_0>0$ and with $A=(A_t)_{t \in [0, T]}$ and $C=(C_t)_{t \in [0, T]}$ adapted processes, with $C_t>0$ a.s. for all $t \in [0,T]$,  .
\item $\Lambda=(\Lambda_t)_{t \in [0, T]}$ satisfies the dynamics
\begin{equation}\notag
d\Lambda_t = \mu'(\Lambda_t)dt+\sigma'(\Lambda_t)dB^4_t, \quad 0 \le t \le T \label{dL},
\end{equation}
$\Lambda_0 \in (\lambda,1)$, with $\mu'$, $\sigma'$ that satisfy conditions Theorem 7 in Chapter \rm{V}.3 in \cite{ProtterBook}. %, with $\int_0^T|M'_s|ds<\infty$,  
%$\int_0^T|\sigma'_s|^2ds<\infty$. 
 Furthermore $\mu'(\lambda)>0$, $\mu'(1)<0$, $\sigma'(1)=0$, $\sigma'(\lambda)=0$ a.s., so that we obtain $\lambda \le \Lambda_t \le 1$, a.s. for all $t \in [0,T]$.
\item $\pi=(\pi_t)_{t \in [0, T]}$ is bounded, i.e. $|\pi_t| \le  \Pi <\infty$ a.s for all $t \in [0,T]$. 
\item $T$ is a bounded a.s. (possibly by a very large constant) $\mathbb{F}$-stopping time independent of $(B^1,B^2,N)$ such that $\tau<T$ a.s.%$\mathbb{E}_P[T|\cal{F}_t]>\mathbb{E}_P[\tau\vee t|\cal{F}_t]$ for all $t \in [0,T)$. 
\end{enumerate}
\end{assumption}
Notice that we assume $\tau <T$ and $T$ bounded a.s. for the sake of simplicity. 
The following results still hold without these conditions by imposing  some integrability conditions on $T$. For example, it would be sufficient $T<\infty$ a.s., $\mathbb{E}_P[e^T|\cal{F}_t]<\infty$ and $\mathbb{E}_P[T-\tau|\cal{F}_t]>0$ a.s. for $t \in [0,T]$.
%From now on we will always write $\mathbb{E}_P[\cdot|\cal{F}_t]$ to mean the conditional expectation under $P$. 
\begin{remark}\label{nobetainsigma}
Here we exclude that $\sigma$ can depend on $\beta$. However the following results also hold for the case $\sigma_t=\alpha\beta_t$, $t \in [\tau,T]$, $\alpha \in \mathbb{R}$, considered 
in  \cite{JarrowProtter2012} to model the evolution of the bubble given by illiquidity effects.
We refer to \cite{thesis} for more details in this case.
%\\ Notice that we assume $\tau <T$ a.s. for the sake of simplicity. The results hold also for the case $P(\tau<T)>0$, with minimal changes. 
\end{remark}

\begin{proposition}\label{propM}
From the hypothesis on $M$ it follows that $\int_0^T M_s^{\alpha}ds<\infty$ a.s. for all $\alpha \in \mathbb{R}$. 
\end{proposition}
\textit{Proof}. 
Following the same argument as in \cite{MijatovicUrusov2012}, we have that 
\begin{align}
\int_0^T M_s^{\alpha}ds=&\int_0^T \frac{M_s^{\alpha}}{\tilde{\sigma}^2(M_s)}d[M,M]_s=\int_0^{\infty}\frac{x^{\alpha}}{\tilde{\sigma}^2(x)}L_T^xdx,
\end{align}
where $L_T^x$ is the local time at $T$ and the last equality follows by occupation time formula (see for example Corollary 1 in Chapter \rm{IV} of \cite{ProtterBook}). \\ Then the integral is finite since, by the fact that $0<M_s<\infty$ a.s. for each $s \in [0,T]$, we have that the occupation time $L_T^a$ has compact support in $(0,\infty)$. $\quad \Box$.\\ 

From Remark \ref{nobetainsigma} we have that $\beta$ satisfies the SDE
\begin{equation}\notag
 d\beta_t=2\Lambda_tM_t\left[(-k\beta_t+\mu_t)dt+\sigma_tdB_t^2+xW_t^FdN_t\right], \quad \tau \le t < T.
\end{equation}
The bubble takes therefore the following explicit expression:
\begin{align}\label{bubble2}
\beta_t=&\beta_{\tau}e^{-k\int_{\tau}^t \Lambda_sM_s ds}+\int_{\tau}^t \mu_s\Lambda_sM_se^{-k\int_s^t \Lambda_uM_u du}ds+\notag \\ &+
 \int_{\tau}^t \sigma_s\Lambda_sM_se^{-k\int_s^t \Lambda_uM_u du}dB_s^2,  \quad \tau \le t < T.
\end{align}

\section{Flow of equivalent local martingale measures}
Let $\cal{M}_{loc}(W)$ be the space of equivalent local martingale measures for $W$. Given $Q \in  \cal{M}_{loc}(W)$, a $Q$-bubble $\beta^Q$ is defined as 
$$\beta_t^Q=W_t-\mathbb{E}_Q[W_T|\cal{F}_t]$$
in the approach of \cite{JarrowProtter2007} and \cite{JarrowProtter2010}. In particular we have that the bubble introduced in (\ref{bubblenew}) coincides with a $Q$-bubble if and only if
$$W_t^F=\mathbb{E}_{Q}[W_T|\cal{F}_t], \quad t \in [0,T]$$
for some $Q \in  \cal{M}_{loc}(W)$. \\
This is of course not possible in our setting. However we can find a flow $(Q^t)_{t \in [0,T]} \subseteq \cal{M}_{loc}(W)$ such that  
\begin{equation}\label{eqzag}
W_t^F=\mathbb{E}_{Q^t}[W_T|\cal{F}_t]=\mathbb{E}_{Q^t}[W_T^F|\cal{F}_t].
\end{equation}
In this way the bubble described in (\ref{bubblenew}) is the result of the shift in the pricing measure induced by the change in the macro-economic and financial conditions in the market.
\begin{remark}\label{remzag}
Note that (\ref{eqzag}) does not imply that $W^F$ is a martingale under $Q^t$. Eq. (\ref{eqzag}) holds $t$-wise and in general it is not true that 
$$W_s^F=\mathbb{E}_{Q^t}[W_T|\cal{F}_s]$$
for $s \ne t$. Furthermore $Q^t$ is an equivalent local martingale measure for $W$ only on $[0,T)$.
\end{remark}

%pricing measure induced by the trading activities. 
We now explicitly compute a flow $(Q^t)_{t \in [0,T)}\in \cal{M}_{loc}(W)$ justifying the existence of the bubble in (\ref{bubblenew}) from a fundamental point of view. \\
Let $Q \in \cal{M}_{loc}(W)$.
 Then the density process $Z=(Z_t)_{t \in [0, T)}$ of $Q$ with respect to $P$ is given by 
\begin{equation}\notag
Z_t=\frac{dQ}{dP}|_{\cal{F}_t}=\mathcal{E}\left(\int_0^{\cdotp} \alpha_s^1dB_s^1+\int_0^{\cdot} \alpha_s^2dB_s^2+\int_0^{\cdot} \alpha_s^3d\tilde{N}_s+\int_0^{\cdot} \alpha_s^4dB_s^3+\int_0^{\cdot} \alpha_s^5dB_s^4+L_t\right)_t, 
\end{equation} 
$0 \le t < T$, where $\tilde{N}_t=N_t-\int_0^{t\wedge \tau}n_sds$, $t \in [0,T)$,  $L$ is a martingale strongly orthogonal to ($B^1,B^2,B^3,B^4,N$) and the processes $\alpha^i$, $i=1,\dots,5$ are such that for $0 \le s < T$ the following equality holds:
  \begin{equation}\label{2Q}
W_s^F(a+b\alpha^1_s)+2\Lambda_sM_s\left(\mu_s+\sigma_s\alpha_s^{2}-k\beta_s\right)\mathbb{1}_{\{s\ge \tau\}} +2\pi_sxW_s^F\Lambda_sM_s(\alpha_s^3+1)\mathbb{1}_{\{s < \tau\}}=0.
\end{equation}
Since (\ref{2Q}) does not involve $\alpha^4$, $\alpha^5$ or $L$, we put $\alpha^4 \equiv \alpha^5 \equiv L \equiv 0$. \\
We can split (\ref{2Q}) as
\begin{align}\label{2Q1}
&b\alpha^{1}_s=-a-2\pi_sx\Lambda_sM_s(\alpha_s^{3}+1) \quad \text{for } s<\tau \\
\intertext{and}
&b\alpha^{1}_s=-a+\frac{2\Lambda_s M_s}{W_s^F}\left(k\beta_s-\mu_s-\sigma_s\alpha_s^{2}\right) \quad \text{for } s\ge\tau \label{2Q2}.
\end{align}
%Since the fundamental value is usually defined as the conditional expectation at $t$ of the future cash flow that an investor would earn holding the asset up to the final time, we want to find a flow of equivalent martingale measures $Q^t \in \cal{M}_{loc}(W)$ such that 
%Futhermore 
%\begin{equation}\label{eqzag}
%W_t^F=\mathbb{E}_{Q^t}[W_T|\cal{F}_t]=\mathbb{E}_{Q^t}[W_T^F|\cal{F}_t], \quad t \in [0,T].
%\end{equation}
%If we let $B^{\mathcal{Q}_t}=(B^{\mathcal{Q}_t,1},B^{\mathcal{Q}_t,2})$ denotice the $Q^t$-s
%tandard Brownian motion given by 
%\begin{align}
%&B^{\mathcal{Q}_t,1}_s=B_s^1-\int_0^s\alpha_u^{t,1}du, \\
%&B^{\mathcal{Q}_t,2}_s=B_s^2-\int_0^s\alpha_u^{t,2}du, 
%\end{align}
%It is important to remark that this condition does not imply that $W^F$ is a martingale under $Q^t$ since $Q^t$ is also depending on $t$. \\
We look for a flow of the form  
\begin{equation}\label{Zmu}
Z_{t,s}=\frac{dQ^t}{dP}|_{\mathcal{F}_s}=\mathcal{E}\left(\int_0^{\cdot} \alpha_u^{t,1}dB_u^1+\int_0^{\cdot} \alpha_u^{t,2}dB_u^2+\int_0^{\cdot} \alpha_u^{t,3}d\tilde{N}_u\right)_s, \quad s \in [0,T),
\end{equation}
since (\ref{2Q}) does not involve conditions on $\alpha^{t,4}$, $\alpha^{t,5}$ and $\alpha^{t,6}$. In particular, we note that the laws of $M$, $\Lambda$ and $T$ are invariant under this 
change of measure. \\
If $\alpha^{t,1}$, $\alpha^{t,2}$ and $\alpha^{t,3}$ satisfy (\ref{2Q1}) and (\ref{2Q2}), the fundamental process under $Q^t$ is given by 
\begin{equation}\label{dWF}
\frac{dW_s^F}{W_s^F}=\tilde{\mu}_s^tds+b d\tilde{B}_s^t, \quad 0 \le s \le T,
\end{equation}
where $\tilde{B}^t$ denote the $Q^t$-standard Brownian motion given by 
\begin{equation}\notag
\tilde{B^t_s}=B_s^1-\int_0^s\alpha_u^{t,1}du, \quad 0 \le s \le T,
\end{equation}
and
\begin{equation}\label{mutilde} 
\tilde{\mu}_s^t= 
\begin{cases}
 -2 \pi_sx\Lambda_sM_s(\alpha_s^{t,3}+1) \quad & \text{for } s < \tau,
  \\
  \frac{2\Lambda_s M_s}{W_s^F}\left(k\beta_s-\mu_s-\sigma_s\alpha_s^{t,2}\right) \quad & \text{for } s \ge \tau. 
\end{cases} 
\end{equation}
If the condition 
\begin{equation}\label{condfin}
\mathbb{E}_{Q^t}\left[\int_t^T(W_s^F)^2 ds\right]<\infty
\end{equation}
 is satisfied, we have that (\ref{eqzag}) is equivalent to 
$$
\mathbb{E}_{Q^t}\left[\int_t^TW_s^F\tilde{\mu}_s^t ds\Big|\cal{F}_t\right]=0,$$
that is
\begin{align}
 0=\mathbb{E}_{Q^t}&\left[ \int_t^{\tau} W_s^F \pi_sx\Lambda_sM_s(\alpha_s^{t,3}+1)ds+\int_{\tau}^T \Lambda_s M_s\left(k\beta_s-\mu_s-\sigma_s\alpha_s^{t,2}\right)ds\Big|\cal{F}_t\right] \label{2mu1}
\end{align}
for $t<\tau$ and
\begin{equation}\label{2mu2}
 \mathbb{E}_{Q^t}\left[\int_t^T \Lambda_s M_s\left(k\beta_s-\mu_s-\sigma_s\alpha_s^{t,2}\right)ds\Big|\cal{F}_t\right]=0   
\end{equation}
for $t \ge \tau$.\\
To show the existence of the flow $(Q^t)_{t \in [0,T)} \subseteq \cal{M}_{loc}(W)$, we choose $\alpha^{t,2}$ and $\alpha^{t,3}$ so that the integrals inside the conditional expectation in (\ref{2mu1}) and (\ref{2mu2}) are zero almost surely. We show later on that a posteriori this choice ensures as well that (\ref{condfin}) holds.\\
For $t \ge \tau$, let 
\begin{equation}\notag
\alpha_s^{t,2}=\frac{1}{\Lambda_s M_s \sigma_s}\left(s-\frac{\mathbb{E}[T|\cal{F}_t]+t}{2}+\frac{\mathbb{E}^2[T|\cal{F}_t]-\mathbb{E}[T^2|\cal{F}_t]}{2(\mathbb{E}[T|\cal{F}_t]-t)}\right)+\frac{k\beta_s}{\sigma_s}-\frac{\mu_s}{\sigma_s}, \quad t \le s < T.
\end{equation} 
Notice that such $\alpha_s^{t,2}$ is well defined since from Assumption \ref{remM} it holds $\Lambda_s > 0$, $M_s > 0$, $\sigma_s>0$ a.s. for every $s \in [0,T]$. \\
With this choice we have on $\{T > t\}$ that 
\begin{align}\notag
&\mathbb{E}_{\mathbb{Q}^t}\left[\int_t^T \Lambda_s M_s\left(k\beta_s-\mu_s-\sigma_s\alpha_s^{t,2}\right)ds\Big|\cal{F}_t\right]\notag\\
=&\mathbb{E}_{\mathbb{Q}^t}\left[\int_t^T\left(s-\frac{\mathbb{E}[T|\cal{F}_t]+t}{2}+\frac{\mathbb{E}^2[T|\cal{F}_t]-\mathbb{E}[T^2|\cal{F}_t]}{2(\mathbb{E}[T|\cal{F}_t]-t)}\right)ds\Big|\cal{F}_t\right] \notag \\
=&\mathbb{E}_{\mathbb{Q}^t}\left[\left(\frac{T^2-t^2}{2}-(T-t)\frac{\mathbb{E}[T|\cal{F}_t]+t}{2}+(T-t)\frac{\mathbb{E}^2[T|\cal{F}_t]-\mathbb{E}[T^2|\cal{F}_t]}{2(\mathbb{E}[T|\cal{F}_t]-t)}\right)\Big|\cal{F}_t\right] \notag \\
=&\frac{\mathbb{E}[T^2|\cal{F}_t]-t^2}{2}-\frac{(\mathbb{E}[T|\cal{F}_t]-t)(\mathbb{E}[T|\cal{F}_t]+t)}{2}+\frac{\mathbb{E}^2[T|\cal{F}_t]-\mathbb{E}[T^2|\cal{F}_t]}{2}=0,
\end{align}
since by Assumption \ref{remM} the law of $T$ does not change under $Q^t$. \\
For $t < \tau$ define 
\begin{equation}\notag
C_{t,\tau}:=\int_t^{\tau} W_s^F \pi_sx\Lambda_sM_s(\alpha_s^{t,3}+1)ds
\end{equation}
and choose $\alpha^{t,2}_s$ to be such that \\% that is to the values of $\alpha^{t,2}_s$ for $s\ge \tau $. \\
$$\mathbb{E}_{Q^t}\left[\int_{\tau}^T \Lambda_s M_s\left(k\beta_s-\mu_s-\sigma_s\alpha_s^{t,2}\right)ds\Big|\cal{F}_t\right]=-\mathbb{E}_{Q^t}\left[C_{t,\tau}|\cal{F}_t\right],$$ 
i.e.
 \begin{align}\notag
\alpha_s^{t,2}=&\frac{1}{\Lambda_s M_s \sigma_s}\left(s-\frac{\mathbb{E}_{Q^t}[C_{t,\tau}|\cal{F}_t]}{\mathbb{E}[T-\tau|\cal{F}_t]}-\frac{\mathbb{E}[T+\tau|\cal{F}_t]}{2}+\frac{\mathbb{E}^2[T|\cal{F}_t]-\mathbb{E}[T^2|\cal{F}_t]}{2\mathbb{E}[T-\tau|\cal{F}_t]}-\frac{\mathbb{E}^2[\tau|\cal{F}_t]-\mathbb{E}[\tau^2|\cal{F}_t]}{2\mathbb{E}[T-\tau|\cal{F}_t]}\right)\notag \\&\quad+\frac{k\beta_s}{\sigma_s}-\frac{\mu_s}{\sigma_s} \notag, \quad t \le s \le T, 
\end{align} 
so that 
\begin{align}\notag
&\mathbb{E}_{\mathbb{Q}^t}\left[\int_{\tau}^T \Lambda_s M_s\left(k\beta_s-\mu_s-\sigma_s\alpha_s^{t,2}\right)ds\Big|\cal{F}_t\right]\notag\\
=&\mathbb{E}_{\mathbb{Q}^t}\left[\int_{\tau}^T\left(s-\frac{\mathbb{E}_{Q^t}[C_{t,\tau}|\cal{F}_t]}{\mathbb{E}[T-\tau|\cal{F}_t]}-\frac{\mathbb{E}[T+\tau|\cal{F}_t]}{2}+\frac{\mathbb{E}^2[T|\cal{F}_t]-\mathbb{E}[T^2|\cal{F}_t]}{2\mathbb{E}[T-\tau|\cal{F}_t]}-\frac{\mathbb{E}^2[\tau|\cal{F}_t]-\mathbb{E}[\tau^2|\cal{F}_t]}{2\mathbb{E}[T-\tau|\cal{F}_t]}\right)ds\Big|\cal{F}_t\right] \notag \\
=&\frac{\mathbb{E}[T^2-\tau^2|\cal{F}_t]}{2}-\mathbb{E}_{Q^t}[C_{t,\tau}|\cal{F}_t]-\frac{\mathbb{E}[T-\tau|\cal{F}_t]\mathbb{E}[T+\tau|\cal{F}_t]}{2}+\frac{\mathbb{E}^2[T|\cal{F}_t]-\mathbb{E}[T^2|\cal{F}_t]}{2}\notag \\ & \quad -\frac{\mathbb{E}^2[\tau|\cal{F}_t]-\mathbb{E}[\tau^2|\cal{F}_t]}{2}\notag \\
=&-\mathbb{E}_{Q^t}[C_{t,\tau}|\cal{F}_t], \notag
\end{align}
and then (\ref{2mu1}) holds.\\
%The choice of $\alpha^{t,3}$ is free: we can take advantage on this by choosing $\alpha^{t,3}$ so that, under some assumptions on $\mu_s$ and 
%$\sigma_s$ specified in the following Section, $Z_{t,T}$ in (\ref{ZtT}) will turn out to be a proper martingale. \\ Choosing 
For $s < t \vee \tau$ we set $\alpha_s^{t,2}=0$.\\
Summarizing: 
\begin{align}&\alpha_s^{t,2}=
\begin{cases}
 0 \quad & \text{for }  s<\tau \vee t,\\
\frac{1}{\Lambda_s M_s \sigma_s}\left(s-\eta_{t,\tau}\right)+\frac{k\beta_s}{\sigma_s}-\frac{\mu_s}{\sigma_s} \quad & \text{for } s\ge\tau \vee t,
\end{cases} \label{alphamu2} \\
\intertext{where}
\eta_{t,\tau}=&\frac{\mathbb{E}_{Q^t}[\int_{t\wedge \tau}^{\tau} W_s^F \pi_sx\Lambda_sM_s(\alpha_s^{t,3}+1)ds|\cal{F}_t]}{2\mathbb{E}[T-\tau\vee t|\cal{F}_t]}-\frac{\mathbb{E}[T+\tau\vee t|\cal{F}_t]}{2}+\frac{\mathbb{E}^2[T|\cal{F}_t]-\mathbb{E}[T^2|\cal{F}_t]}{2\mathbb{E}[T-\tau \vee t|\cal{F}_t]}\notag \\ &-\frac{\mathbb{E}^2[\tau\vee t|\cal{F}_t]-\mathbb{E}[(\tau\vee t)^2|\cal{F}_t]}{2\mathbb{E}[T-\tau \vee t|\cal{F}_t]} \label{etamu}.
\end{align}
\begin{remark}\label{remeta}
Notice that from Assumption \ref{remM} and from the fact that the integral in (\ref{etamu}) is bounded, we have that $\eta_{t,\tau}$ is finite and $\cal{F}_t$-measurable, and that moreover $\mathbb{E}[\eta_{t,\tau}^{\alpha}]<\infty$ for all $\alpha \in \mathbb{R}$.
\end{remark}
Choosing
\begin{align}
\alpha^{t,3}_s=
\begin{cases} 
0 & \quad  \text{for } s < t \text{ or } s \ge \tau,\\
\frac{1}{(M_s+1)(W_s^F+1)}-1&\quad  \text{for } t\le s <\tau,
\end{cases} \label{alphamu3}
\end{align}
and 
\begin{align}
&\alpha_s^{t,1}=
\begin{cases}
0 & \quad  \text{for } s < t,\\
-\frac{a}{b}-\frac{2}{b}\pi_s\Lambda_s\frac{M_s}{M_s+1}\frac{1}{W_s^F+1} & \quad  \text{for } t \le s < \tau, 
\\-\frac{a}{b}-\frac{2}{b W_s^F}\left(s-\eta_{t,\tau}\right) & \quad \text{for } s\ge\tau \vee t.
\end{cases} \label{alphamu1}
\end{align}
we have that (\ref{2mu1}) and (\ref{2mu2}) hold. \\
\\ Now we give the following
\begin{proposition}
Let $\alpha^{t,1}$, $\alpha^{t,2}$ and $\alpha^{t,3}$ be as in (\ref{alphamu2})-(\ref{alphamu1}). Then 
 $$\mathbb{E}_{Q^t}\left[\int_t^T(W_s^F)^2 ds\right]<\infty, \quad t \in [0,T].
$$
\end{proposition}
\textit{Proof.} 
From (\ref{mutilde}) and from the expressions of  $\alpha^{t,1}$, $\alpha^{t,2}$ and $\alpha^{t,3}$ in (\ref{alphamu2})-(\ref{alphamu1}) we have that
\begin{equation}
\tilde{\mu}_s^t= 
\begin{cases}
 -2 \pi_sx\Lambda_s\frac{M_s}{M_s+1}\frac{1}{W_s^F+1} \quad & \text{for } s < \tau,
  \\
 \frac{1}{W_s^F}(\eta_{t,\tau}-s) \quad & \text{for } s \ge \tau,
\end{cases} \notag
\end{equation} 
where $\eta_{t,\tau}$ is given in (\ref{etamu}). Then from (\ref{dWF}) it holds that under $Q^t$
\begin{align}
dW_s^F=&  \psi_sds+bW_s^Fd\tilde{B}_s^t\quad & \text{for } s < \tau, \notag \\
dW_s^F=&  (\eta_{t,\tau}-s)ds+bW_s^Fd\tilde{B^t_s}\quad & \text{for } s \ge \tau, \notag
\end{align} 
where $\psi_s=-2\pi_sx\Lambda_s\frac{M_s}{M_s+1}\frac{1}{W_s^F+1}$. \\Thus we have
\begin{equation}
W_s^F=\begin{cases}
 e^{b\tilde{B}_s^t-\frac{b^2}{2}s}\int_0^s\psi_u e^{-b\tilde{B}_u^t+\frac{b^2}{2}u} du\quad & \text{for } s < \tau,
  \\
   e^{b\tilde{B}_s^t-\frac{b^2}{2}s}\int_0^s (\eta_{t,\tau}-u) e^{-b\tilde{B}_u^t+\frac{b^2}{2}u} du\quad & \text{for } s \ge \tau.
\end{cases} \notag
\end{equation} 
Then
\begin{align}
&\mathbb{E}_{Q^t}\left[\int_t^T(W_s^F)^2 ds\right] \notag \\
=&\mathbb{E}_{Q^t}\left[\int_{t \wedge \tau}^{\tau} \left( \int_0^s \psi_u e^{b(\tilde{B}_s^t-\tilde{B}_u^t)-\frac{b^2}{2}(s-u)}du \right)^2 ds+\int_{\tau}^T \left( \int_0^s (\eta_{t,\tau}-u)e^{b(\tilde{B}_s^t-\tilde{B}_u^t)-\frac{b^2}{2}(s-u)}du \right)^2  ds\right] \notag \\
\le &\mathbb{E}_{Q^t}\left[4\Pi^2x^2\int_{t \wedge \tau}^{\tau} \left( \int_0^s e^{b(\tilde{B}_s^t-\tilde{B}_u^t)-\frac{b^2}{2}(s-u)}du \right)^2 ds+(|\eta_{t,\tau}|+T)^2\int_{\tau}^T \left( \int_0^s e^{b(\tilde{B}_s^t-\tilde{B}_u^t)-\frac{b^2}{2}(s-u)}du \right)^2  ds\right] \notag \\
\le &\left(4\Pi^2x^2+\mathbb{E}\left[(|\eta_{t,\tau}|+T)^4\right]^{\frac{1}{2}}\right)\mathbb{E}_{Q^t}\left[\int_{t \wedge \tau}^{T} \left( \int_0^s e^{b(\tilde{B}_s^t-\tilde{B}_u^t)-\frac{b^2}{2}(s-u)}du \right)^4 ds\right]^{\frac{1}{2}} \notag.
\end{align}
Since $T$ is bounded and the first term is finite by Remark \ref{remeta}, it remains to prove 
\begin{equation}\label{int4fin}
\mathbb{E}_{Q^t}\left[\int_{t \wedge \tau}^{T} \left( \int_0^s e^{b(\tilde{B}_s^t-\tilde{B}_u^t)-\frac{b^2}{2}(s-u)}du \right)^4 ds\right]<\infty.
\end{equation}
We have that
\begin{align}
&\mathbb{E}_{Q^t}\left[\int_{t \wedge \tau}^{T} \left( \int_0^s e^{b(\tilde{B}_s^t-\tilde{B}_u^t)-\frac{b^2}{2}(s-u)}du \right)^4 ds\right]  = \mathbb{E}_{Q^t}\left[\int_{t \wedge \tau}^{T} \left( \int_0^s e^{b\tilde{B}_{s-u}^t-\frac{b^2}{2}(s-u)}du \right)^4 ds\right] \notag \\
& = \mathbb{E}_{Q^t}\left[\int_{t \wedge \tau}^{T} \left( \int_0^s e^{b\tilde{B}_r^t-\frac{b^2}{2}r}dr \right)^4 ds\right] \le 
\mathbb{E}_{\mathbb{Q}^t}\left[(T-t\wedge \tau)^2\right]^{\frac{1}{2}} \mathbb{E}_{Q^t}\left[\left( \int_0^T e^{b\tilde{B}_r^t-\frac{b^2}{2}r}dr \right)^8 \right]^{\frac{1}{2}}. \notag
\end{align}
The first term is finite by Assumption \ref{remM} on $T$ and $\tau$, whereas
\begin{align}
&\mathbb{E}_{Q^t}\left[\left( \int_0^T e^{b\tilde{B}_r^t-\frac{b^2}{2}r}dr \right)^8\right] \le \mathbb{E}_{Q^t}\left[ \int_0^T e^{8b\tilde{B}_r^t-4b^2r}dr\right] = \int_0^T \mathbb{E}_{Q^t}\left[e^{8b\tilde{B}_r^t-4b^2r} \right] dr <\infty.\notag
\end{align} 
Then (\ref{int4fin}) holds and we have the result. $\quad \square$ \\
We have therefore proved that, if we take $\alpha^{t,1}$, $\alpha^{t,2}$ and $\alpha^{t,3}$ as in (\ref{alphamu2})-(\ref{alphamu1}), then 
 (\ref{2Q1}), (\ref{2Q2}) and (\ref{eqzag}) are satisfied. \\% (\ref{2Q1}), (\ref{2Q2}) and (\ref{eqzag}) are satisfied for 
%\begin{equation}\label{Zmu}
%Z_{t,T}=\frac{dQ^t}{dP}|_{\mathcal{F}_T}=\mathcal{E}\left(\int_0^T \alpha_s^{t,1}dB_s^1+\int_0^T \alpha_s^{t,2}dB_s^2+\int_0^T \alpha_s^{t,3}d\tilde{N}_s\right). 
%\end{equation}
From now on we denote $Z_{t,s}:=\frac{dQ^t}{dP}|_{\cal{F}_s}$ for all $s \ge t$, and $Z_{t,s}=1$ for $s < t$. \\
Note that we have not yet used the hypothesis on $\mu$ and $\sigma$ of Assumption \ref{remM} to derive (\ref{Zmu}). 
From now on we will need them to prove that $(Z_{t,s})_{s \in [t,T]}$ is a true martingale for each $t \in [0,T]$, i.e. that each $Q^t$, $t \in [0,T]$, in (\ref{Zmu}) belongs to $\in \cal{M}_{loc}(W)$.

\begin{remark}
By Assumption \ref{remM}, as proved in Proposition \ref{propM}, we exclude that the integral $\int_0^{\cdot} M_s^2ds$ can explode in finite time. This is a difference with respect to \cite{JarrowProtter2012}, where the bubble bursts (i.e. $\beta_t=0$) at $\inf\{ s \big| \int_0^s M_u^2du=+\infty$\}. \\
In our model, however, the bubble can be zero, and also negative, even if the liquidity is not zero: by (\ref{bubble2}) it can be seen that this can happen when the drift $\mu$ of the aggregate trading volume becomes negative. In this approach, therefore, whether or not the bubble is positive depends more on the attitude of the investors than on the liquidity. In Section \ref{secnet} we propose an example to show how contagion between traders in financial networks can determine the value of $\mu$. 
\end{remark}

%\subsection{Novikov's condition}
%Summing up, taking into account also the assumptions on the processes $M$, $\Lambda$ and $\pi$ defined in the previous Section, we take a setting explicated by the following
%\begin{Assumption}\label{hypo}
%The process $M=$
%\end{Assumption}

% We want to prove that, taking the processes $M$, $\lambda$ and $\pi$ as specified in Assumption \ref{remM}, $\alpha^{t,1}$, $\alpha^{t,2}$ and $\alpha^{t,3}$ in (\ref{alphamu3})-(\ref{alphamu2}) with such $\mu_t$ and $\sigma_t$ are such that, for all fixed $t \ge 0$, $(Z_{t,s})_{s \in [0,T]}$ with
%\begin{equation}\label{Ztre}
%Z_{t,s}=\mathcal{E}\left(\int_0^s \alpha_u^{t,1}dB_u^1+\int_0^s \alpha_u^{t,2}dB_u^2+\int_0^T \alpha_s^{t,3}d\tilde{N}_s\right)
%\end{equation}
%is a martingale (with respect to time $s$), so that $Q^t$ in (\ref{Zmu}) is equivalent to $P$ for all $t$. From now on, we fix $t$. \\
From now on, we fix $t\in [0,T]$. 
We begin the analysis by noticing that, since $[B^1,N] \equiv [B^2,N]\equiv 0$,
\begin{align}
Z_{t,s}=& \mathcal{E}\left(\int_0^s \alpha_u^{t,1}dB_u^1+\int_0^s \alpha_u^{t,2}dB_u^2+\int_0^s \alpha_u^{t,3}d\tilde{N}_u\right)\notag\\=&\mathcal{E}\left(\int_0^s \alpha_u^{t,1}dB_u^1+\int_0^s \alpha_u^{t,2}dB_u^2\right)\mathcal{E}\left(\int_0^s \alpha_u^{t,3}d\tilde{N}_s\right)
 \notag
\end{align}
for $s \in [0,T]$.\\
Moreover
\begin{align}
\mathcal{E}\left(\int_0^s \alpha_u^{t,3}d\tilde{N}_u\right)\le&\exp \left \{\int_0^s \left[\alpha_u^{t,3}-\frac{1}{2}(\alpha^{t,3}_u)^2\right]dN_u- \int_0^s \alpha_u^{t,3}\pi_udu \right\} \cdot\notag \\& \cdot \prod_{0 \le u \le s}(1+\Delta (\alpha_u^{t,3}N_u))\exp\{\Delta (\alpha_u^{t,3}N_u)+\frac{1}{2}\Delta(\alpha_u^{t,3}N_u)^2\} \notag\\ 
\le & 2\exp \left \{\frac{3}{2}+\int_0^s \left[|\alpha_u^{t,3}|+\frac{1}{2}|\alpha^{t,3}_u|^2\right]dN_u+ \int_0^s |\alpha_u^{t,3}|\pi_udu \right\} \notag \\ 
\le & 2e^{3+T\Pi} \notag,
\end{align}
since by (\ref{alphamu3}) it holds $|\alpha^{t,3}_s| \le 1$. \\
Then, taking $(\bar{Z}_{t,s})_{s \in [0,T]}$ with
\begin{equation}\notag
\bar{Z}_{t,s}=\mathcal{E}\left(\int_0^s \alpha_u^{t,1}dB_u^1+\int_0^s \alpha_u^{t,2}dB_u^2\right)
\end{equation}
we have
\begin{equation}\label{ZlessbarZ}
Z_{t,s}\le 2e^{3+T\Pi}\bar{Z}_{t,s}. %\le C \mathbb{E}\left[\mathcal{E}\left(\int_0^s \alpha_u^{t,1}dB_u^1+\int_0^s \alpha_u^{t,2}dB_u^2\right)\right],
\end{equation}
%since $T$ is bounded by Assumption \ref{remM}. \\
We give the following
\begin{lemma}\label{LemmaDL}
Let $X$, $Y$ be two positive stochastic processes such that $Y_t \le X_t$ a.s. $\forall t \ge 0$, and let $X$ be of class $DL$\footnote{A stochastic process $X$ is of class $DL$ if, for each $t \ge 0$, $\{X_{\tau}: \tau \le t \text{ stopping time}\}$ is uniformly integrable.}. Then $Y$ is of class $DL$ as well. 
\end{lemma}
\textit{Proof.}  By Theorem 11 of Chapter \rm{I} of \cite{ProtterBook} we have that a family of random variables $(U_{\alpha})_{\alpha \in A}$ is uniformly integrable if and only if there exists a function $G$ defined on $[0,\infty)$, positive, increasing and convex, such that $\lim_{x \to \infty}\frac{G(x)}{x}=+ \infty$ and 
$\sup_{\alpha}\mathbb{E}[G \circ |U_\alpha|]<\infty$. Fix now $t\ge 0$, and call $J_t=\{\tau: \tau \le t  \text{ stopping time}\}$,
  $U_X^t=\{X_{\tau}: \tau \in J_t\}$ and $U_Y^t=\{Y_{\tau}: \tau\in J_t\}$. \\
Since by hypothesis $U_X^t$ is uniformly integrable, there exists a function $G$ that satisfies the properties stated before. We have that
$$G(Y_{\tau}) \le G(X_{\tau}), \quad a.s. \text{ for } \tau \in J_t, $$
and then that
$$\mathbb{E}[G(Y_{\tau})] \le \mathbb{E}[G(X_{\tau})], \quad \tau \in J_t. $$
Thus 
$$\sup_{\tau \in J_t} \mathbb{E}[G(Y_{\tau})] \le \sup_{\tau \in J_t} \mathbb{E}[G(X_{\tau})]<\infty.$$
Therefore $U_Y^t$ is uniformly integrable and $Y$ is of class $DL$. $\Box$ \\
 We have then the following
\begin{proposition}\label{3to2}
$(Z_{t,s})_{s \in [0,T]}$ in (\ref{Zmu}) is a martingale if $(\bar{Z}_{t,s})_{s \in [0,T]}$
is a martingale.
\end{proposition}
\textit{Proof.}  Since a local martingale is a true martingale if and only if it is of class $DL$, see Proposition 1.7 of Chapter \rm{IV} of \cite{RevuzYor}, we have that if $\bar{Z}$ is a true martingale then $2e^{3+T\Pi}\bar{Z}$, being a martingale as well, is of class $DL$. Thus, by Lemma \ref{LemmaDL} and by (\ref{ZlessbarZ}), $Z$ is of class $DL$, and therefore by Proposition 1.7 of Chapter \rm{IV} of \cite{RevuzYor} it is a true martingale. $\Box$

  %THESIS
  
%  From now on, therefore, we will check the martingale property for $(\bar{Z}_{t,s})_{s \in [0,T]}$ in (\ref{Zdue}).\\
%We note that the Novikov condition is not satisfied since for example the integrand $\alpha_s^{t,1}$ contains the term $\frac{1}{W^F}$ with $$W_s^F=\exp\left(\left(\mu-\sigma^2/2\right)s+\sigma B_s^1\right),$$ and it can be seen that the expectation of
%the double exponential of the Brownian motion under $P$ is not finite. The same problem holds for the term $\frac{1}{\beta}$ with $\beta$ in (\ref{bubble}). \\
%Therefore, since the other terms are strictly positive and bounded, we can not rely on Novikov condition.

%\begin{remark}
%What we want to prove is that $\mathbb{E}[Z_{t,T}]=1$, where, since $\alpha^{t,1}_u$ and $\alpha^{t,2}_u$ are zero for $u<t$, we can write
%\begin{equation}
%Z_{t,T}=\mathcal{E}\left(\int_t^T \alpha_u^{t,1}dB_u^1+\int_t^T \alpha_u^{t,2}dB_u^2\right)
%\end{equation}
%As we will see later, since the term $\eta_{t,\tau}$ defined in (\ref{eta}) and appearing in (\ref{al1}) and (\ref{al2}) can be estimated with a term that is $\cal{F}_t$-measurable, it will be easier to prove
%\begin{equation}\label{expectt}
%\mathbb{E}[Z_{t,T}|\cal{F}_t]=1.
%\end{equation}
%Then it will easily follow $\mathbb{E}[Z_{t,T}]=\mathbb{E}[\mathbb{E}[Z_{t,T}|\cal{F}_t]]=1$.
%\end{remark}

%To prove that (\ref{expectt}) holds 

To prove that $\bar{Z}$ is a martingale we rely on some results provided by Mijatovic and Urusov \cite{MijatovicUrusov} and by Wong and Heyde \cite{WongHeyde}. We first need some preliminaries.
\\ Consider the state space $J=(l,r)$, $-\infty\le l < r \le \infty$ and a $J$-valued diffusion $Y=(Y_s)_{s \in [0,T]}$ on some filtered probability space, governed by the SDE
\begin{equation}\label{ymij}
dY_s=\mu_Y(Y_s)ds+\sigma_Y(Y_s)dB_s, \quad 0 \le s \le T,
\end{equation}
with $Y_0=x_0 \in J$, $W$ Brownian motion and deterministic functions $\mu_Y(\cdot)$ and $\sigma_Y(\cdot)$, that from now on we will simply denote by $\mu_Y$ and $\sigma_Y$, such that 
\begin{equation}\label{condsigma1}
\sigma_Y(x) \ne 0 \quad \forall x \in J
\end{equation}
 and 
 \begin{equation}\label{condsigma2}
 \frac{1}{\sigma_Y^2},\text{ }  \frac{\mu_Y}{\sigma_Y^2} \in L_{loc}^1(J),
 \end{equation}
 where $ L_{loc}^1(J)$ denotes the class of locally integrable functions on $J$, i.e. the measurable functions $(J,\mathcal{B}(J)) \rightarrow (\mathbb{R},\mathcal{B}(\mathbb{R}))$ that are integrable on compact subsets of $J$. \\
% From now on we will denote $\mu_Y=\mu_Y(\cdot)$, $\sigma_Y=\sigma_Y(\cdot)$.. \\ 
Consider the stochastic exponential
\begin{equation}\label{stocexp}
\cal{E}\left(\int_0^s f(Y_u)dB_u\right), \quad 0 \le s \le T,
\end{equation}
with $f(\cdot)$ such that 
\begin{equation}\label{condb1}
\frac{f^2}{\sigma_Y^2} \in L_{loc}^1(J)
\end{equation}
 and the auxiliary $J$-valued diffusion $\tilde{Y}$ governed by the SDE
\begin{equation}\label{Ytilde}
d\tilde{Y}_s=\left(\mu_Y(\tilde{Y}_s)+f(\tilde{Y}_s)\sigma_Y(\tilde{Y}_s)\right)ds+\sigma_Y(\tilde{Y}_s)d\tilde{B}_s, \quad 0 \le s \le T,
\end{equation}
where $\tilde{B}$ is a Brownian motion on some probability space $(\tilde{\Omega},\tilde{\cal{F}},\tilde{P})$. \\
Put $\bar{J}=[l,r]$ and, fixing an arbitrary $c \in J$, define
\begin{align}
&\rho(x):=\exp\left\{-\int_c^x\frac{2\mu_Y}{\sigma_Y^2}(y)dy\right\}, \quad x \in J, \label{rho} \\
&\tilde{\rho}(x):=\rho(x)\exp\left\{-\int_c^x\frac{2f}{\sigma_Y}(y)dy\right\}, \quad x \in J, \label{rhotilde} \\
&s(x):=\int_c^x\rho(y)dy, \quad x \in \bar{J}, \label{s} \\
& \tilde{s}(x):=\int_c^x\tilde{\rho}(y)dy,\quad x \in \bar{J}. \label{stilde}
\end{align}
Denote $\rho=\rho(\cdot)$, $s=s(\cdot)$, $s(r)=\lim_{x \to r^-}s(x)$,  $s(l)=\lim_{x \to l^+}s(x)$, and analogously for $\tilde{s}(\cdot)$ and $\tilde{\rho}(\cdot)$.\\
Recall that by Feller's test for explosions $\tilde{Y}$ exits its state space with positive probability at the boundary point $r$ if and only if 
\begin{equation}\label{cond1}
\tilde{s}(r)<\infty \quad \text{and} \quad \frac{\tilde{s}(r)-\tilde{s}}{\tilde{\rho}\sigma_Y^2} \in L_{loc}^1(r-),
\end{equation}
where $L_{loc}^1(r-):=\{g | g:(J,\mathcal{B}(J)) \rightarrow (\mathbb{R},\mathcal{B}(\mathbb{R})) \text{ such that } \int_x^r ~g(y)dy<\infty \text{ for some } x \in J\}$. 
%denote the class of Borel functions $f:J \rightarrow \mathbb{R}$ such that $\int_x^r|f(y)| dy<\infty$ for some $x \in J$. \\
Similarly, $\tilde{Y}$ exits its state space with positive probability at the boundary point $l$ if and only if 
\begin{equation}\label{cond2}
\tilde{s}(l)>-\infty \quad \text{and} \quad \frac{\tilde{s}-\tilde{s}(l)}{\tilde{\rho}\sigma_Y^2} \in L_{loc}^1(l+),
\end{equation}
where $L_{loc}^1(l+):=\{g | g:(J,\mathcal{B}(J)) \rightarrow (\mathbb{R},\mathcal{B}(\mathbb{R})) \text{ such that } \int_l^x ~g(y)dy<\infty \text{ for some } x \in J\}$ 
%denote the class of Borel functions $f:J \rightarrow \mathbb{R}$ such that $\int_l^x|f(y)| dy<\infty$ for some $x \in J$. \\
Moreover, the endpoint $r$ of $J$ is said to be \textit{good} if  
\begin{equation}\label{cond3}
s(r)<\infty \quad \text{and} \quad \frac{(s(r)-s)f^2}{\rho\sigma_Y^2} \in L_{loc}^1(r-),
\end{equation}
or equivalently (see \cite{MijatovicUrusov}) if 
\begin{equation}\label{cond4}
\tilde{s}(r)<\infty \quad \text{and} \quad \frac{(\tilde{s}(r)-\tilde{s})f^2}{\tilde{\rho}\sigma_Y^2} \in L_{loc}^1(r-).
\end{equation}
Similarly, the endpoint $l$ of $J$ is said to be \textit{good} if  
\begin{equation}\label{cond5}
s(l)>-\infty \quad \text{and} \quad \frac{(s-s(l))f^2}{\rho\sigma_Y^2} \in L_{loc}^1(l+),
\end{equation}
or equivalently if 
\begin{equation}\label{cond6}
\tilde{s}(l)>-\infty \quad \text{and} \quad \frac{(\tilde{s}-\tilde{s}(l))f^2}{\tilde{\rho}\sigma_Y^2} \in L_{loc}^1(l+).
\end{equation}
We recall here Theorem 2.1 in \cite{MijatovicUrusov}.

\begin{theorem}\label{uru}
Let the functions $\mu_Y$, $\sigma_Y$, and $f$ satisfy conditions (\ref{condsigma1}), (\ref{condsigma2}) and  (\ref{condb1}), and let $Y$ be a solution of the SDE (\ref{ymij}). 
\\ Then the Dol\'eans exponential given by (\ref{stocexp}) is a martingale for any $T<\infty$ if and only if both of the following requirements are satisfied:
\begin{enumerate}[(a)]
\item condition (\ref{cond1}) does not hold or conditions (\ref{cond3})-(\ref{cond4}) hold;
\item condition (\ref{cond2}) does not hold or conditions (\ref{cond5})-(\ref{cond6}) hold.
\end{enumerate}  
\end{theorem}

We now obtain the following

 \begin{proposition}\label{lemmamart}
 Let $S=(S_s)_{s \in [0,T]}$ be a geometric Brownian motion
 \begin{equation}\label{geombr}
dS_s=\mu_0 S_sds+\sigma_0 S_s dB_s, \quad 0 \le s \le T,
\end{equation} 
where $B$ is a Brownian motion, $\mu_0 \in \mathbb{R}$ and $\sigma_0>0$. \\
Then the process $$\mathcal{E}\left(\int_0^{s}(S_u)^{-1}dB_u\right), \quad 0 \le s \le T,$$ is a martingale.
 \end{proposition}

\textit{Proof}. We show that the requirements of Theorem \ref{uru} hold for $Y=S$, with $\mu_Y(x)=\mu_0x$, $\sigma_Y(x)=\sigma_0x$ and $f(x)=x^{-1}$. Notice that $\mu_Y$, $\sigma_Y$ and $f$ satisfy conditions (\ref{condsigma1}), (\ref{condsigma2}) and  (\ref{condb1}) with $J=(0,\infty)$.  Then, taking $c=1$ for the functions (\ref{rho})-(\ref{stilde}) and first assuming $\frac{2\mu_0}{\sigma^2_0} \ne 1$, we have
\begin{align}
& \rho(x)=\exp\left\{-\int_1^x\frac{2\mu_Y}{\sigma_Y^2}(y)dy\right\}=x^{\frac{-2\mu_0}{\sigma_0^2}}, \label{rholemma} \\
 &\tilde{\rho}(x)=\rho(x)\exp\left\{-\int_1^x\frac{2f}{\sigma_Y}(y)dy\right\}=x^{\frac{-2\mu_0}{\sigma_0^2}}\exp\left({\frac{2}{\sigma_0}\left(\frac{1}{x}-1\right)}\right),  \label{rhotildelemma} \\
&s(x)=\int_1^x\rho(y)dy=\frac{\sigma_0^2}{2\mu_0-\sigma_0^2}(1-x^{-\gamma_0}),  \label{slemma} \\
& \tilde{s}(x)=\int_1^x\tilde{\rho}(y)dy=e^{-\frac{2}{\sigma_0}}\left(-\frac{2}{\sigma_0}\right)^{-\gamma_0}\left[\bar{\Gamma}\left(\gamma_0, -\frac{2}{x\sigma_0}\right)-\bar{\Gamma}\left(\gamma_0, -\frac{2}{\sigma_0}\right)\right], \label{stildeex}
\end{align}
with $\gamma_0=\frac{2\mu_0}{\sigma_0^2}-1$ and where $\bar{\Gamma}(a,z)=\int_z^{\infty} e^{-t} t^{a-1}dt$, $a \in \mathbb{R}^+$, $z \in \mathbb{R}$, is the incomplete Gamma function extended to all $\mathbb{R}$. \\
Notice that in (\ref{stildeex}) we have that
\begin{align}\notag
\tilde{s}(x)&=e^{-\frac{2}{\sigma_0}}\left(-\frac{2}{\sigma_0}\right)^{-\gamma_0}\left[\bar{\Gamma}\left(\gamma_0, -\frac{2}{x\sigma_0}\right)-\bar{\Gamma}\left(\gamma_0, -\frac{2}{\sigma_0}\right)\right] \notag \\ 
&=e^{-\frac{2}{\sigma_0}}\left(\frac{2}{\sigma_0}\right)^{-\gamma_0}(-1)^{-\gamma_0}  \int_{-\frac{2}{x\sigma_0}}^{-\frac{2}{\sigma_0}} e^{-t}(-1)^{\gamma_0-1} |t|^{\gamma_0-1}dt  \notag \\ &=
-e^{-\frac{2}{\sigma_0}}\left(\frac{2}{\sigma_0}\right)^{-\gamma_0} \int_{-\frac{2}{x\sigma_0}}^{-\frac{2}{\sigma_0}} e^{-t} |t|^{\gamma_0-1}dt \in \mathbb{R}.
\end{align}

% that has the following asymptotic behavior for $z \in \mathbb{R}$ (see for example \cite{Abramowitz}):
%\begin{itemize}
%\item if $a>0$, $\lim_{z \to 0} \Gamma(a,z)=\Gamma(a)>0$, whereas if $a<0$, $\Gamma(a,z) \sim -\frac{z^a}{a} $ for $z \to 0$ (from above or from below, so when $z \to 0$ from below the limit can be negative);
%\item $\Gamma(a,z) \sim e^{-z} z^{s-1}$ for $z \to \infty$;
%\item $\lim_{z\to -\infty} y^{-a}\left(\Gamma(a,z)-\Gamma(a,y)\right)=-\infty$, with $a\in \mathbb{R}$, $y \le 0$ .
%\end{itemize}
%Notice that in (\ref{stildeex}) we have real outcomes since in general $y^{-a}\left(\Gamma(a,z)-\Gamma(a,y)\right)$ is real when $sgn(z)=sgn(y)$, as it is in our case. \\
We obtain that:

\begin{itemize}
\item in $l=0$ we have $$\tilde{s}(0) =-e^{-\frac{2}{\sigma_0}}\left(\frac{2}{\sigma_0}\right)^{-\gamma_0} \int_{-\infty}^{-\frac{2}{\sigma_0}} e^{-t} |t|^{\gamma_0-1}dt =-\infty,$$ thus condition (\ref{cond2}) does not hold and the first requirement of (b) in Theorem \ref{uru} is fulfilled;
\item  if $\gamma_0<0$ we have $$\tilde{s}(\infty)=e^{-\frac{2}{\sigma_0}}\left(\frac{2}{\sigma_0}\right)^{-\gamma_0} \int_{-\frac{2}{\sigma_0}}^0 e^{-t} |t|^{\gamma_0-1}dt = \infty
$$ then condition (\ref{cond1}) does not hold and the first requirement of (a) in Theorem \ref{uru} is fulfilled;
\item if $\gamma_0>0$ then $s(\infty)=\frac{\sigma_0^2}{2\mu_0-\sigma_0^2}=C<\infty$, and condition (\ref{cond3}) holds since $$\frac{s(r)-s}{\rho\sigma_0^2}=C\frac{x^{-\gamma_0}x^{\frac{2\mu_0}{\sigma_0^2}}}{x^4}=\frac{1}{x^3}.$$ Therefore the second requirement of (a) in Theorem \ref{uru} is fulfilled.
\end{itemize}
So we have that if $\gamma_0 \ne 0$ the requirements of Theorem \ref{uru} are satisfied, and thus $Z$ is a martingale. \\
In the case $\gamma_0=0$, i.e. $\mu_0=\frac{\sigma_0^2}{2}$, we have that the process $S=(S_u)_{u \in [0,T]}$ in (\ref{geombr}) takes the form $S_u=e^{\sigma_0 B_u}$, $0 \le u \le T$. We can thus apply the results of Theorem \ref{uru} taking $J=(-\infty,\infty)$, $\mu_Y \equiv 0$, $\sigma_Y \equiv 1$, $f(x)=e^{-\sigma_0 x}$ and $c=0$ in (\ref{rho})-(\ref{stilde}). We have 
\begin{align}
& \rho(x)=\exp\left\{-\int_0^x\frac{2\mu_Y}{\sigma_Y^2}(y)dy\right\}=1,  \notag \\
 &\tilde{\rho}(x)=\rho(x)\exp\left\{-\int_0^x\frac{2f}{\sigma_Y}(y)dy\right\}=\exp\left(2(e^{-\sigma_0 x}-1)/\sigma_0\right), \\ \notag
&s(x)=\int_0^x\rho(y)dy=x  \\ \notag
& \tilde{s}(x)=\int_0^x\tilde{\rho}(y)dy=\frac{1}{\sigma_0}e^{-\frac{2}{\sigma_0}}\left(Ei\left(2/\sigma_0\right)-Ei\left(2e^{-\sigma_0 x}/\sigma_0\right)\right),  \notag
\end{align}
where $Ei(z)=-\int_{-z}^{\infty}\frac{e^{-u}}{u}du$ is the exponential integral function that satisfies $\lim_{z\to \infty}Ei(z)=\infty$ and $\lim_{z\to 0}Ei(z)=-\infty$. Therefore $\tilde{s}(\infty)=\infty$ and $\tilde{s}(-\infty)=-\infty$, then the first requirements of (a) and (b) of Theorem \ref{uru} are both satisfied and $Z$ is a martingale. $\quad \Box$
Then we have immediately 
\begin{corollary}\label{corWF}
%Take  the filtration  $\mathbb{F}=(\cal{G}_s)_{s \in \mathbb{R}^+}$ generated by ($B^1,B^2,B^3,B^4,N$) that we consider in Section \ref{def}, and let $\beta$ be as in (\ref{beta}), 
%with $M$ and $\Lambda$ bounded and driven my Brownian motions $B^3$ and $B^4$ independent of $B^2$.  
Under Assumptions \ref{remM}, the process 
\begin{equation}\label{EWsF}
\cal{E}\left(\int_{\tau}^s \frac{1}{W^F_u}dB_u^1\right), \quad \tau \le s  \le T,
\end{equation}
is a martingale for every fixed $T<\infty$.
\end{corollary}

To prove that Corollary \ref{corWF} also implies that $\cal{E}\left(\int_{\tau}^s \alpha_u^{t,1}dB_u^1\right)$ is a martingale, we extend the results of Wong and Heyde in \cite{WongHeyde}.\\
To this purpose we consider a $\mathbb{F}$-progressively measurable $d$-dimensional process $H=(H_s)_{s \in [0,T]}$ of the form
\begin{equation}\label{XWongHeyde}
H_s=\xi(B(\cdot),s)\zeta_s+\eta_s,
\end{equation}
where $\xi\in C_0(\mathbb{R}^{d+1}, \mathbb{R}^d)$, $B$ is a $d$-dimensional progressively measurable Brownian motion and $\zeta, \eta$ are $d$-dimensional stochastic processes independent of $B$. Here the product between $\xi$ and $\zeta$ is intended componentwise.  \\ 
Define  
\begin{equation}\notag
\tau_N^{M_H}=\inf\left(s \in [0,T]: M_H(t):= \int_0^t \|H_u \|^2 du \ge N \right),
\end{equation}
 with the convention that $\inf \emptyset =\infty$, and then 
 \begin{equation}\label{tauwong}
\tau^{M_H}=\lim_{N \to \infty}\tau_N^{M_H}.
\end{equation}
 Then we have the following
 
 \begin{proposition}\label{propwonghayde}
 Let $H$ be as in (\ref{XWongHeyde}), and defined up to the explosion time  $\tau^{M_H}$ in (\ref{tauwong}). 
 Then there also exists a $d$-dimensional $\mathbb{F}$-progressively measurable process, $Y=(Y_s)_{s \in [0,T]}$ with $Y_s=\xi(W(\cdot)+\int_0^{\cdot}Y_udu ,s)\zeta_s+\eta_s$ defined up to the explosion time  $\tau^{M_Y}$ with
 \begin{equation}\notag
  \tau^{M_Y}=\lim_{N \to \infty}\tau_N^{M_Y},
 \end{equation}
 where
 \begin{equation}\notag
\tau_N^{M_Y}=\inf\left(s \ge 0: M_Y(s):= \int_0^s \|Y_u \|^2 du \ge N \right) \wedge T,
\end{equation}
such that the stochastic exponential $Z^H=(Z^H_s)_{s \in[0,T]}$ with $Z^H_s=\cal{E}\left(\int_0^s H_u dW_u \right)$ satisfies 
\begin{equation}\notag
P(\tau^{M_Y}>T)=\mathbb{E}[Z^H_T].
\end{equation}
Hence $Z^H$ is a (true) martingale if and only if $P(\tau^{M_Y}>T)=1$.
\end{proposition}
% The proof for the case $\zeta_s=1$ in (\ref{XWongHeyde}) can be found in \cite{WongHeyde}, but it can be checked that the results can be extended to a general $\zeta$ in (\ref{XWongHeyde}), if the process $\zeta$ does not depend on the Brownian motion $W$. 
\textit{Proof.} Since the proof is a long but easy extension of the result in \cite{WongHeyde}, we omit it here and refer to \cite{thesis}. $\quad \Box$
 \begin{proposition}\label{propalpha1}
 In the setting of Section \ref{def}, the process 
 \begin{equation}\notag
\mathcal{E}\left(\int_0^s |\alpha_u^{t,1}|dB_u^1\right), \quad 0 \le s \le T,
\end{equation}
with $\alpha^{t,1}$ in (\ref{alphamu1})  is a martingale for each $t \in [0,T]$. 
\end{proposition}
\textit{Proof}. 
For $s<\tau$ we have 
\begin{equation}\notag
|\alpha_s^{t,1}| =\frac{a}{b} + \frac{2}{b}\pi_s\Lambda_s\frac{M_s}{M_s+1}\frac{1}{W_s^F+1} \le \frac{a}{b}+\frac{2}{b}\Pi,
\end{equation}
then $\mathcal{E}\left(\int_0^{\cdot} |\alpha_u^{t,1}|dB_u^1\right)$ is a martingale up to time $\tau$ since it satisfies Novikov condition since 
$$
\mathbb{E}\left[\exp\left(\int_0^{\tau}(\alpha_s^{t,1})^2ds\right)\right]\le \mathbb{E}\left[\exp(c^2\tau)\right]
$$
with $c=\frac{a}{b}+\frac{2}{b}\Pi$.\\
Consider now $s \ge \tau$. We have that the process $Y$ associated to $|\alpha_s^{t,1}|$ as in Proposition \ref{propwonghayde} satisfies
 \begin{equation}\notag
Y_s=\frac{2}{bW_s^F}(s+|\eta_{t,\tau}|)\exp\left(-b\int_{t \wedge \tau}^sY_udu\right), \quad t\wedge \tau \le s \le T,
\end{equation}
with $\eta_{t,\tau}$ in (\ref{etamu}). On the other hand, we have
 \begin{equation}\notag
\tilde{Y}_s=\frac{1}{W_s^F}\exp\left(-b\int_{t \wedge \tau}^s\tilde{Y}_udu\right), \quad t\wedge \tau \le s \le T,
\end{equation} 
where $\tilde{Y}$ is the process associated to $\frac{1}{W^F}$. By Corollary \ref{corWF} and Proposition \ref{propwonghayde} it holds 
\begin{equation}\label{Ytildenotinf}
\int_{t \wedge \tau}^T \tilde{Y}_s^2 < \infty.
\end{equation}
We want to see that the integral of $Y^2$ does not explode as well. \\
We have that 
\begin{equation}\label{Delta}
\Delta_s=\frac{\tilde{Y}_s}{Y_s}=\frac{b}{s+|\eta_{t,\tau}|}\cdot \exp\left(b\int_{t \wedge \tau}^s(Y_u-\tilde{Y}_u)du\right), \quad  t \wedge \tau \le s \le T. 
\end{equation}
\\ Define the stopping time 
\begin{equation}\notag
 \tau_1=\inf\{s \in [ t \wedge \tau,T] : \Delta_s \le 1\}\wedge T
\end{equation}
and notice that, since $Y$ and $\tilde{Y}$ are continuous, $\Delta_{\tau_1}=1$. \\ Define 
\begin{equation}\notag
 \tau_2=\inf\{s \ge \tau_1 : \Delta_s \ge 1\}\wedge T,
\end{equation}
If $\tau_1=T$, we are done. Otherwise consider $s \in (\tau_1,\tau_2).$ \\ 
Since for $\tau_1 < s < \tau_2$ we have
\begin{align}\notag
&\Delta_s=\frac{b}{s+|\eta_{t,\tau}|}\cdot \exp\left(b\int_{t\wedge \tau}^s(Y_u-\tilde{Y}_u)du\right)
% &=\frac{\tilde{\zeta}_s}{\zeta_s} \cdot \exp\left(\int_0^{\tau_1}\psi_s(Y_s-\tilde{Y}_s)ds\right) \cdot \exp\left(\int_{\tau_1}^u\psi_s(Y_s-\tilde{Y}_s)ds\right) \\
 \ge \frac{b}{T+|\eta_{t,\tau}|} \exp\left(b\int_{t \wedge \tau}^{\tau_1}(Y_u-\tilde{Y}_u)du\right),
 \end{align}
 it follows
\begin{equation}\notag
Y_s \le \frac{\tilde{Y}_s\left(T+|\eta_{t,\tau}|\right)}{b}  \exp\left(b\int_{t \wedge \tau}^{\tau_1}(\tilde{Y}_u-Y_u)du\right)  \le \frac{\tilde{Y}_s\left(T+|\eta_{t,\tau}|\right)}{b}  \exp\left(b\int_{t \wedge \tau}^{\tau_1}\tilde{Y}_udu\right)
\end{equation}
for $\tau_1 < s < \tau_2$, which implies, together with (\ref{Ytildenotinf}), that $M_Y(s):=\int_{t \wedge \tau}^sY_s^2ds$ does not explode before $\tau_2$. \\ But after $\tau_2$, up to $\tau_3=\inf\{s \ge \tau_2 : \Delta_s \le 1\} \wedge T$, $Y$ is 
smaller than $\tilde{Y}$, hence $M_Y(s) \le M_{\tilde{Y_s}}$ on $[\tau_2,\tau_3]$. \\
Repeating this argument up to $T$, we obtain that $\mathcal{E}\left(\int_0^s |\alpha_u^{t,1}|dB_u^1\right)$ is a martingale by Proposition \ref{propwonghayde}. $\quad \Box$ \\
%
%
%
%Taking first $s<\tau$ we have 
%\begin{equation}\notag
%|\alpha_s^{t,1}| =\frac{a}{b} + \frac{2}{b}\pi_s\Lambda_s\frac{M_s}{M_s+1} \le \frac{a}{b}+\frac{2}{b}\Pi,
%\end{equation}
%and then that $\alpha_s^{t,1}$ is bounded a.s.\\
%For $s \ge \tau$ we have instead 
% \begin{equation}\notag
%|\alpha_s^{t,1}| \le 
%\frac{a}{b}+\frac{2}{b W_s^F}\left|\frac{|\eta_{t,\tau}|+T}{2}+T\right| 
%\end{equation}
%with 
%\begin{equation}\label{esteta}
%|\eta_{t,\tau}| \le W_t^F\ \frac{exp\{2\int_{t\wedge\tau}^{\tau}\pi_u x\Lambda_u \frac{M_u}{M_u+1} du\}+1}{T-
%\tau}+\tau\vee t \le W_t^F \frac{e^{2xT\Pi}+1}{T-\tau}+T.
%\end{equation}
%Hence we can write
%\begin{equation}\notag
%|\alpha^{t,1}_s| \le \frac{K_{t,1}}{W_s^F}+\frac{a}{b}
%\end{equation}
%with $$K_{t,1}=\frac{2}{b}\left(W_t^F \frac{e^{2xT\Pi}+1}{T-\tau}+2T\right)<\infty$$ and $\cal{F}_t$-measurable.
 We want now to prove that 
 \begin{equation}\label{propconv}
\mathcal{E}\left(\int_0^s |\alpha_u^{t,2}|dB_u^2\right), \quad 0 \le s \le T,
\end{equation}
with $\alpha^{t,2}$ in (\ref{alphamu2})  is a martingale as well.\\
We start with the following
\begin{proposition}\label{propbetaint}
Let $\beta$ be the bubble as in (\ref{bubble2}). Under Assumption \ref{remM}, the Dol\'eans exponential 
 \begin{equation}\notag
\mathcal{E}\left(\int_0^s \beta_udB^2_u\right), \quad 0 \le s \le T,
\end{equation}
is a martingale.
\end{proposition}
\textit{Proof}. If we rewrite $\beta$ in the form (\ref{XWongHeyde}), we obtain that
$$\xi(B^2(\cdot),s)=\int_{\tau}^s \sigma_u\Lambda_uM_ue^{-k\int_u^s k\Lambda_rM_r dr}dB_u^2, \quad \tau \le s \le t,$$
i.e. the process $Y$ associated to $\beta$ in Proposition \ref{propwonghayde} is given by
 \begin{align}\notag
Y_s=&\beta_{\tau}e^{\int_{\tau}^s (-k+\sigma_u)\Lambda_uM_u ds}+\int_{\tau}^s \mu_u\Lambda_uM_ue^{\int_u^s (-k+\sigma_r)\Lambda_rM_r dr}du\notag \\ &
+ \int_{\tau}^s \sigma_u\Lambda_uM_ue^{\int_u^s (-k+\sigma_r)\Lambda_rM_r dr}dB_u^2,  \quad \tau \le s \le T. \label{intY}
\end{align}
We first prove that $Y_s<\infty$ for each $s \in [\tau,T]$. We have $\int_u^s (-k+\sigma_r)\Lambda_rM_r dr<\infty$ a.s. for each $s \in [\tau, T]$ by the hypothesis on $\sigma$ and $\Lambda$ in Assumption \ref{remM} and by Proposition \ref{propM}. \\ 
Thus by Theorem 2.4 of \cite{MijatovicUrusov2012} and by the fact that $T$ is bounded, we obtain
\begin{equation}\label{intMnotinfty}
\int_{\tau}^T e^{\alpha\int_u^s (-k+\sigma_r)\Lambda_rM_r dr}du<\infty
\end{equation}
for all $\alpha \in \mathbb{R}$, and then by the hypothesis on $\mu$ in Assumption \ref{remM}, and again by Proposition \ref{propM}, we have 
$$\int_{\tau}^s \mu_u\Lambda_uM_ue^{\int_u^s (-k+\sigma_r)\Lambda_rM_r dr}du<\infty,\quad \tau \le s \le T.$$ 
By (\ref{intMnotinfty}) and by Assumption \ref{remM} it follows that the stochastic integral in (\ref{intY}) does not explode before $T$, so we have that $Y_s<\infty$ for each $s \in [\tau,T]$. \\ We prove that this implies $\int_{t \wedge \tau}^TY_s^2ds<\infty$.  By the expression of $Y$ in (\ref{intY}) we have
\begin{align}
\int_{t \wedge \tau}^TY_s^2ds =  &\int_{t \wedge \tau}^TY_s^2\frac{1}{M_s^2\Lambda_s^2\sigma_s^2}d[Y,Y]_s \notag \\ 
\intertext{(by the Kunita-Watanabe inequality)}
 \le & \left(\int_{t \wedge \tau}^TY_s^4d[Y,Y]_s\right)^{1/2}\left(\int_{t \wedge \tau}^T\frac{1}{M_s^4\Lambda_s^4\sigma_s^4}d[Y,Y]_s\right)^{1/2} \notag \\
\intertext{(by the occupation time formula)}
= & \left(\int_{-\infty}^{\infty}a^4L_T^ada\right)^{1/2}\left(\int_{t \wedge \tau}^T\frac{1}{M^2_s\Lambda^2_s\sigma^2_s}ds\right)^{1/2}<\infty:
\end{align}
 the first integral is finite because  the local time $L_T^a$ has bounded support in $(-\infty,\infty)$, since $Y$ does not explode before $T$, and the second one
is finite by Assumption \ref{remM} and Proposition \ref{propM}. Then the result follows by Proposition \ref{propwonghayde}. $\quad \Box$
\\
%We start with the following
%\begin{proposition}
%Let $V=(V_s)_{s \in [0,T]}$ be a process driven by the SDE 
%\begin{equation}\notag
%dV_s=U_s\left[(\varphi_sV_s+\psi_s)ds+\theta_sdB_s\right], \quad 0 \le s \le T,
%\end{equation}
%with $B$ Brownian motion, $U$, $\varphi$, $\psi$ progressively measurable processes such that $|U_s|, |\varphi_s|, | \psi_s| <\infty$ a.s. for each $s \in [0,T]$, $0<\theta_s<\infty$ a.s. for each $s \in [0,T]$. Then the process
% \begin{equation}\notag
%\mathcal{E}\left(\int_0^s V_udB_u\right), \quad 0 \le s \le T,
%\end{equation}
%is a martingale.
%\end{proposition}
%\textit{Proof}. The process $V$ has the form 
%\begin{align}\notag
%V_s=&V_0 e^{\int_0^sU_u\varphi_u du}+\int_0^s U_u\psi_ue^{\int_{u}^s U_r\varphi_r dr}du+\int_0^t U_u\theta_ue^{\int_{u}^s U_r\varphi_r dr}dB_u,  \quad 0 \le s \le T,
%\end{align}
%so that it can be easily seen that the process $Y$ associated to $V$ in Proposition \ref{propwonghayde} is
%\begin{align}\notag
%Y_s=&Y_0 e^{\int_0^sU_u(\varphi_u+\theta_u) du}+\int_0^s U_u\psi_ue^{\int_{u}^s U_r(\varphi_r+\theta_r) dr}du+\int_0^s U_u\theta_ue^{\int_{u}^s U_r(\varphi_r+\theta_r) dr}dB_u, \quad 0 \le s \le T.
%\end{align}
%Let's prove first that 
%
%Hence by Proposition \ref{propwonghayde} the result follows. $\quad \Box$ \\
\begin{proposition}\label{propalpha2}
Under Assumption \ref{remM} the process 
\begin{equation}\notag
\mathcal{E}\left(\int_0^s |\alpha_u^{t,2}|dB_u^2\right), \quad 0 \le s \le T,
\end{equation}
with $\alpha^{t,2}$ in (\ref{alphamu2})  is a martingale for each $t \in [0,T]$.
\end{proposition}
\textit{Proof.} 
We have that
\begin{align}\notag
|\alpha_s^{t,2}| & \le \frac{1}{\sigma_s}\left(\frac{\eta_{t,\tau}+T}{\lambda M_s}+k\mu_s+k|\beta_s|\right), \quad \tau \wedge t \le s \le T.
%\frac{1}{\alpha\Lambda_s\beta_sM_s}\left|T+\frac{\eta_{t,\tau}+T}{2}\right|+\frac{\mu_s}{\alpha\beta_s}+\frac{k}{\alpha}\\& \le
%\frac{1}{\alpha\lambda\beta_sM_s}\left(W_t^F \frac{e^{2xT\Pi}+1}{T-\tau}+2T\right)+\frac{\bar{M}}{\alpha\beta_s}+\frac{k}{\alpha}
\end{align}
%where
%$$\Phi_s=\frac{c}{M_s}+k\mu_s, \quad \tau \le s \le T,$$
Let $\tilde{Y}$ be the process associated to $\frac{\eta_{t,\tau}+T}{\lambda M_s}+k\mu_s+k|\beta|$ in Proposition \ref{propwonghayde}, and $\bar{Y}$ the one associated to $k|\beta|$. \\ We have
\begin{align}
\tilde{Y}_s=&\frac{\eta_{t,\tau}+T}{\lambda M_s}+k\mu_s+k|\beta_s|+k\int_{\tau}^s\sigma_u\Lambda_uM_u\tilde{Y}_ue^{-k\int_u^s\Lambda_rM_rdr}du \notag \\
=& \frac{\eta_{t,\tau}+T}{\lambda M_s}+k\mu_s+k|\beta_s|+k\int_{\tau}^s\sigma_u\Lambda_uM_u\bar{Y}_ue^{-k\int_u^s\Lambda_rM_rdr}du\notag \\ &\quad+k\int_{\tau}^s\sigma_u\Lambda_uM_u(\tilde{Y}_u-\bar{Y}_u)e^{-k\int_u^s\Lambda_rM_rdr}du \notag \\
=&\frac{\eta_{t,\tau}+T}{\lambda M_s}+k\mu_s+\bar{Y}_s+k\int_{\tau}^s\sigma_u\Lambda_uM_u(\tilde{Y}_u-\bar{Y}_u)e^{-k\int_u^s\Lambda_rM_rdr}du, \quad \tau \le s \le T, \notag
\end{align}
and consequently, for $\bar{D}_s:=\tilde{Y}_s-\bar{Y}_s$,  
%\begin{equation}\notag
%\bar{D}_s:=\tilde{Y}_s-\bar{Y}_s=\frac{\eta_{t,\tau}+T}{\lambda M_s}+k\mu_s+k\int_{\tau}^s\sigma_u\Lambda_uM_u\bar{D}_ue^{-k\int_u^s\Lambda_rM_rdr}du, \quad \tau \le s \le T,
%\end{equation} 
%i.e.
\begin{equation}\notag
d\bar{D}_s=d\left(\frac{\eta_{t,\tau}+T}{\lambda M_s}+k\mu_s\right)+k\Lambda_sM_s\left[(\sigma_s-1)\bar{D}_s+\frac{\eta_{t,\tau}+T}{\lambda M_s}+k\mu_s\right]ds, \quad \tau \le s \le T,
\end{equation} 
so that we can write
\begin{align}\notag
\bar{D}_s%=&\frac{c}{M_s}+k\mu_s+k\int_{\tau}^s\Phi_u\sigma_u\Lambda_uM_ue^{k\int_u^s\Lambda_rM_r(\sigma_r-1)dr}du \notag \\
=&\frac{\eta_{t,\tau}+T}{\lambda M_s}+k\mu_s+k\int_{\tau}^s\left(\frac{\eta_{t,\tau}+T}{\lambda}+k\mu_uM_u \right)\sigma_u\Lambda_ue^{k\int_u^s\Lambda_rM_r(\sigma_r-1)dr}du     \quad \tau \le s \le T.
\end{align} 
By Assumption \ref{remM} and by Proposition \ref{propM}, with the same argument as in the proof of Proposition \ref{propbetaint}, we have that
$$
\int_{t \wedge \tau}^T \bar{D}_s^2ds=\int_{t \wedge \tau}^T |\tilde{Y}_s-\bar{Y}_s|^2ds<\infty.
$$
Then, since by Proposition \ref{propbetaint} we have $\int_{t \wedge \tau}^T |\bar{Y}_s|^2ds<\infty$, we obtain 
\begin{equation}\label{intYtilde}
\int_{t \wedge \tau}^T |\tilde{Y}_s|^2ds<\infty.
\end{equation}
%and then that $\mathcal{E}\left(\int_{\tau}^s \left(\Phi_s+k\beta_s\right)dB_u^2\right),$ $0 \le s \le T,$ is a martingale. \\
Now call $Y$ the process associated to $R^{t,2}$ in Proposition \ref{propwonghayde}. \\
It holds
\begin{align}
Y_s%=&\frac{1}{\sigma_s}\left(\frac{\eta_{t,\tau}+T}{\lambda M_s}+k\mu_s+k|\beta_s|+k\int_{\tau}^s\Lambda_uM_uY_ue^{-k\int_u^s\Lambda_rM_rdr}du \right)\notag \\
=&\frac{1}{\sigma_s}\left(\frac{\eta_{t,\tau}+T}{\lambda M_s}+k\mu_s+k|\beta_s|+k\int_{\tau}^s\Lambda_uM_u\tilde{Y}_ue^{-k\int_u^s\Lambda_rM_rdr}du \right)\notag \\ & \quad+\frac{1}{\sigma_s}k\int_{\tau}^s\Lambda_uM_u(Y_u-\tilde{Y}_u)e^{-k\int_u^s\Lambda_rM_rdr}du \notag \\
=&\frac{1}{\sigma_s}\left(\tilde{Y}_s+k\int_{\tau}^s\Lambda_uM_u(Y_u-\tilde{Y}_u)e^{-k\int_u^s\Lambda_rM_rdr}du \right)\notag, \quad \tau \le s \le T.
\end{align}
Then we have
\begin{align}
\sigma_s Y_s-\tilde{Y}_s=%& k \int_{\tau}^s\sigma_u\Lambda_uM_u(Y_u-\tilde{Y}_u)e^{-k\int_u^s\Lambda_rM_rdr}du\notag \\
&\Psi_s+k \int_{\tau}^s\Lambda_uM_u(\sigma_uY_u-\tilde{Y}_u)e^{-k\int_u^s\Lambda_rM_rdr}du, \quad \tau \le s \le T, \notag 
\end{align}
where $(\Psi_s)_{s \in [\tau,T]}$ is given by
\begin{equation}\label{Psi}
\Psi_s=k \int_{\tau}^s\Lambda_uM_u(\tilde{Y}_u-\sigma_u\tilde{Y}_u)e^{-k\int_u^s\Lambda_rM_rdr}du, \quad \tau \le s \le T.
\end{equation}
It follows that $D_s=\sigma_sY_s-\tilde{Y}_s$ satisfies
\begin{equation}\notag
 dD_s=d\Psi_s+k\Lambda_sM_s\Psi_sds, \quad \tau \le s \le T,
\end{equation}
and so that it takes the form 
\begin{equation}\notag
 D_s=\Psi_s+k\int_{\tau}^s\Lambda_uM_u\Psi_udu, \quad \tau \le s \le T.
\end{equation}
Since by Assumption \ref{remM} the process $\Psi$ in (\ref{Psi}) does not explode before $T$, $D_s=\sigma_s Y_s-\tilde{Y}_s<\infty$ a.s. for each $s \in [0,T]$. \\
Thus, with the same argument as in the proof of Proposition \ref{propbetaint} it can be proved that
$$
 \int_{t \wedge \tau}^T |\sigma_s Y_s-\tilde{Y}_s|^2ds<\infty.
$$
By (\ref{intYtilde}) we then have
$$
 \int_{t \wedge \tau}^T |\sigma_s Y_s|^2ds<\infty.
$$
Then by the integrability hypothesis on $\frac{1}{\sigma^4}$ in \rm(ii) of Assumption \ref{remM} it holds 
$$
 \int_{t \wedge \tau}^T |Y_s|^2ds<\infty.
$$
 The result then follows by Proposition \ref{propwonghayde} and by the fact that if $Y^{\alpha}$ is the process associated to $|\alpha^{t,2}|$ it can easily seen that $Y_s^{\alpha} \le Y_s$ a.s. for each $s \in [\tau,T]$. $\quad \Box$
\begin{proposition}\label{propositionz1z2}
Consider $(Z_{t,s}^1)_{s \in[0,T]}$ and $(Z_{t,s}^2)_{s \in[0,T]}$, with
\begin{align}
&Z^1_{t,s}=\mathcal{E}\left(\int_0^s \alpha_u^{t,1}dB_u^1\right) \label{Z1}
\intertext{and}
&Z_{t,s}^2=\mathcal{E}\left(\int_0^s \alpha_u^{t,2}dB_u^2\right), \label{Z2}
\end{align}
where $\alpha^{t,1}$ and $\alpha^{t,2}$ are as in (\ref{alphamu1}) and (\ref{alphamu2}), and suppose that Assumption \ref{remM} holds.   
 \\Then $(Z_{t,s}^1)_{s \in[0,T]}$ and $(Z_{t,s}^2)_{s \in[0,T]}$ are true martingales.
%\begin{equation}
%\mathbb{E}[Z^1_{t,T}]=\mathbb{E}[Z^2_{t,T}]=1.
%\end{equation}
\end{proposition}
The proof follows by Proposition \ref{propalpha1}, by Proposition \ref{propalpha2} and by the following
\begin{lemma}\label{lemmaabsval}
Consider $H_s=\int_0^s Y_u dB_u$ and $\bar{H}_s=\int_0^s |Y_u| dB_u$, $s \ge 0$, where $Y$ is a stochastic process such that the stochastic integral is well defined. Then   
$\cal{E}(H)$ is a martingale if and only if $\cal{E}(\bar{H})$ is a martingale.
\end{lemma}
\textit{Proof}.  Theorem 4.1 in \cite{BleiEngelbert} states that, for a general continuous local martingale $H$, $\cal{E}(H)$ is a martingale if and only if 
$$\lim_{n \to \infty} Q_s(\{A_s<n\})=1 \text{ for all } s \ge 0,$$
where $A_s=[H,H]_s$ and $dQ_s=\cal{E}(H_{T_s})dP$, and $T_s:=\inf\{u \ge 0: A_u>s\}$. Since $[H,H]_s=\int_0^s Y^2_u du=\int_0^s |Y_u|^2 du=[\bar{H},\bar{H}]_s$, this property hold for $H$ if and only of it holds for $\bar{H}$. Hence we have the result. $\quad \Box$

We are now ready to state the main result of the Section:
\begin{theorem}\label{maintheo}
Under Assumption \ref{remM}, $Q^t$ defined in (\ref{Zmu}) belongs to $\cal{M}_{loc}(W)$ for each $t \in [0,T)$. 
\end{theorem}
\textit{Proof} 
The proof follows by the fact that taking $\alpha^{t,1}$ and $\alpha^{t,2}$ as in (\ref{alphamu1}) and (\ref{alphamu2}), with $\mu_t$, $\sigma_t$, $M$, $\Lambda$ and $\pi$ satisfying Assumption \ref{remM}, then $(\bar{Z}_{t,s})_{s \in[0,T]}$ with
\begin{equation}\notag
\bar{Z}_{t,s}=\mathcal{E}\left(\int_0^s \alpha_u^{t,1}dB_u^1+\int_0^s \alpha_u^{t,2}dB_u^2\right)
\end{equation}
is a martingale with respect to time $s$.\\
This follows immediately from Proposition \ref{propositionz1z2}: $(Z^1_{t,s})_{s \in[0,T]}$ in (\ref{Z1}) and $(Z^2_{t,s})_{s \in[0,T]}$ in (\ref{Z2})
are martingales, so by Proposition \ref{propwonghayde} we know that $H^1=\alpha^{t,1}$ and $H^2= \alpha^{t,2}$ are such that the associated processes $Y^1$ and $Y^2$ defined in Proposition \ref{propwonghayde} do not explode before $T$. Taking now $H=(H^1,H^2)$, the associated process $Y=(Y^1,Y^2)$ does not explode before $T$ as well, and this concludes the proof. $\quad \Box$

\begin{remark}
Note that Theorem \ref{maintheo} also implies that $\cal{M}_{loc}(W) \neq \emptyset$, hence that our market model is arbitrage-free on $[0,T)$.
\end{remark}

%\begin{remark}
%Notice that, proving the existence of such a flow of martingale measures for $W$, we have also automatically proved that the market is arbitrage-free since $\cal{M}_{loc}(W) \ne \emptyset.$
%\end{remark}

%\begin{example}
%In \cite{JarrowProtter2012} the authors take the aggregate trading volume as given by
%\begin{equation}\notag
%dX_t=\alpha\beta_tdB_t^2, \quad \tau \le t \le T.
%\end{equation}
%This is the case when $\mu_t=0$ for all $t \in [\tau,T]$: the bubble corresponds to $\bar{\beta}$ given in $(\ref{barbubble})$, and taking 
% $\alpha^{t,1}$ and $\alpha^{t,3}$ in (\ref{alphamu1}) and in (\ref{alphamu3}) respectively and 
% $\alpha^{t,2}$ in (\ref{alphamu2}) given by
%\begin{equation}
%\alpha_s^{t,2}=
%\begin{cases}
% 0 \quad & \text{for }  s<\tau \vee t\\
%\frac{1}{\Lambda_s M_s \alpha\beta_s}\left(s-\frac{\eta_{t,\tau}+T}{2}\right) \quad & \text{for } s\ge\tau \vee t \notag
%\end{cases} 
%\end{equation}
%with $\eta_{t,\tau}$ in (\ref{etamu}) we have that
%\begin{equation}\notag
%Z_{t,T}=\frac{dQ^t}{dP}|_{\mathcal{F}_T}=\mathcal{E}\left(\int_0^T \alpha_s^{t,1}dB_s^1+\int_0^T \alpha_s^{t,2}dB_s^2+\int_0^T \alpha_s^{t,3}d\tilde{N}_s\right)
%\end{equation}
%is a martingale and that $Q^t$ is equivalent to $P$ for all $t \in [0,T]$. 
%\end{example}

\section{Liquidity induced bubbles in a network}\label{secnet}

As an illustration of the previous results, we focus on a particular example. We note however that the results of this section are of independent interest since we provide one of the few contributions on mathematical modeling of bubbles in a network. For further results on this topic, we also refer to \cite{Battiston}, where it is shown how bubbles can have an impact on the structure of a banking network, and to \cite{Bouchard}, where the authors describe the passage from a well-connected network with high global confidence to a poorly connected network with low global confidence, producing a boom and bust cycle. Our approach is however quite different: we consider a network of $N$ investors who may be influenced by the trading activity of their neighborhoods. Investors may place a buy market order on the bubbly asset because their neighborhoods in the network have bought the asset as well. 
We model the trading contagion mechanism between agents taking place from time $\tau$ via the evolution dynamics of the aggregate trading volume. 
%We call $X^k=(X^k_t)_{\tau \le t \le T}$ the stochastic adapted process standing for the aggregate trading volume generated by the trading of one general investor of degree $k$ in the network, and we let
Our analysis is based on some epidemiological studies, which describe how diseases spread in social networks, or how computer viruses spread from computer to computer. In particular, we focus on the SIS model, studied for example by Pastor-Satorras and Vespignani (see \cite{PastorVespignani2001} and \cite{SatorrasVespignani2001}) to analyze virus diffusion in a population.\\
The aggregate trading volume of an investor of degree $k$ in the network is given by the adapted stochastic process $X^k=(X_t^k)_{t \in [\tau,T]}$. 
 Put
 \begin{equation}\label{exprho}
 \bar{X}^k_t=\mathbb{E}[X^k_t], \quad \tau \le t < T.
 \end{equation}  

We assume that 
\begin{equation}\label{xkbar}
\frac{d\bar{X}^k_t}{dt}  =-\delta \bar{X}^k_t +  \lambda k b_t \Theta(\bar{\rho}_t)  (\theta_t-\bar{X}^k_t),  \quad \tau \le t < T,
\end{equation}
where $\bar{\rho}$ is a continuous function representing the expected fraction of investors who are holding the asset (i.e. that have bought the asset and not already sold it), $\Theta(\bar{\rho}_t)$ is the probability that an individual at the end of an edge has done a trade before or at time $t$, $\lambda$ is the rate of trading contagion and $\delta$ is the rate of selling. Furthermore $b$ and $\theta$  are continuous functions standing for the medium amount of asset traded per buyer and the medium amount of wealth of the investors, respectively.\\% (here we suppose that all investors have the same amount of wealth). \\
Now we focus on the expression of $ \Theta(\bar{\rho}_t)$. As stated by Pastor-Satorras and Vespignani \cite{PastorVespignani2001}, one could be tempted to impose $\Theta(x)=x$, but this approximation can be too strong for networks with an highly inhomogeneous density, for example for networks with a power-law degree distribution.  \\ In particular, by Bayes rule and since for any given node $v$ it holds $$P(\text{meet }v|deg(v)=k)=\frac{k}{\sum_j jq_j}$$ where $q_j$ is the number of nodes with degree $j$, we have that 
\begin{equation}\notag
P(deg(v)=k|\text{meet }v)=\frac{P(\text{meet }v|deg(v)=k)P(deg(v)=k)}{P(\text{meet }v)}=\frac{k}{\frac{1}{N}\sum_j jq_j}p_k=\frac{kp_k}{z}.
\end{equation}
%where $N$ is the total number of nodes.\\
Therefore, as pointed out in \cite{GaiKapadia} and \cite{NewmanStrogatzWatts}, we have 
\begin{equation}\label{theta}
\Theta(\bar{\rho}_t)=\frac{1}{z}\sum_k kp_k\bar{\rho}^k_t,  \quad \tau \le t < T,
\end{equation}
where $\bar{\rho}^k=(\bar{\rho}^k_t)_{t\in [\tau,T)}$ is the expected fraction of investors of degree $k$ that are holding the asset. Notice that, if the degree distribution is very peaked at the average degree $z$ so that we can approximate $p_{\left \lfloor{z} \right \rfloor}\approx 1$, $p_k\approx 0$ for $k \ne \left \lfloor{z} \right \rfloor$, then
$$\Theta(\bar{\rho}_t)\approx \bar{\rho}^{\left \lfloor{z} \right \rfloor}_t \approx \bar{\rho}_t.$$
Since $\bar{X}_t^k=b_t\bar{\rho}_t^k$, substituting (\ref{theta}) in (\ref{xkbar}) we find 
\begin{equation}\notag
\frac{d\bar{X}^k_t}{dt}  =-\delta \bar{X}^k_t + \lambda \frac{\sum_j jp_j\bar{X}^j_t }{z}k(\theta_t-\bar{X}^k_t),  \quad \tau \le t < T,
\end{equation}
and thus, considering $\bar{X}_t=\sum_k q_k\bar{X}_t^k$, where $q_k$ is the number of investors of degree $k$,  
\begin{equation}\label{assxkbar}
\frac{d\bar{X}_t}{dt}  =-\delta\bar{X}^k_t + \lambda N \bar{n}_t\left(\theta_t-\frac{\bar{n}_t}{z}\right),  \quad \tau \le t < T,
\end{equation}
where $N$ is the number of investors in the network and
\begin{equation}\label{net}
\bar{n}_t=\sum_k k p_k \bar{X}^k_t,  \quad \tau \le t < T.
\end{equation} 
Since (\ref{exprho}) holds, we may assume that the aggregate trading volume satisfies  
\begin{equation}\label{Xours1}
dX_t=\left(-\delta X_t +\lambda N n_t\left(\theta_t-\frac{n_t}{z}\right)\right)dt+\bar{\sigma}_tX_t^{\alpha}(N\theta_t-X_t)^{\alpha}dB_t^2, \quad \tau \le t < T,
\end{equation}
where the drift is induced by (\ref{assxkbar}). Here 
\begin{equation}\label{nXk}
n_t=\sum_k k p_k X^k_t, \quad \tau \le t < T,
\end{equation}
 $\alpha>1/2$, $\lambda>\delta$, and $\bar{\sigma}=(\bar{\sigma_t})_{t \in [\tau,T)}$ is a progressively measurable process such that 
 \begin{equation}\label{eqbarsigma}
 \int_{\tau}^T\bar{\sigma}_s^4ds<\infty
 \end{equation}
 and $\int_{\tau}^T\frac{1}{\bar{\sigma}_s^4}ds<\infty$. Furthermore we assume
\begin{equation}\label{inttheta}
\int_{\tau}^T \theta_s^{8\alpha}ds<\infty.
\end{equation}
 Notice that from (\ref{nXk}) and since $X_t^k \le \theta_t$ a.s. for all $k$ and for all $t \in [\tau,T)$ it follows $\frac{n_t}{z}\le \theta_t$ and $X_t=\sum_k q_k X_t^k \le N\theta_t$  a.s. for all $t \in [\tau,T)$.\\
%a progressively measurable process  at time $t$, such that $\int_{\tau}^Tb_s^2ds<\infty$,
%When the drift of (\ref{Xours1}) becomes negative, the sellers begin to be more than the buyers, and the bubble start bursting: we restrict our attention to positive drifts, i.e. only to the building phase of the bubble, since the
%mechanisms after the burst of the bubble are different from the ones during the increase of the bubble. 
We are in the framework of Section \ref{def}, with 
\begin{equation}\label{munet}
\mu_t=-\delta X_t +\lambda N n_t\left(\theta_t-\frac{n_t}{z}\right)
\end{equation}
 and 
 \begin{equation}\label{sigmanet}
 \sigma_t=\bar{\sigma}_tX_t^{\alpha}(N\theta_t-X_t)^{\alpha}.
\end{equation}
We have the following SDE for the bubble $\beta$:
\begin{equation}\label{ourbubble}
d\beta_t=\Lambda_tM_t\left[-k\beta_t+2\left(-\delta X_t +\lambda N n_t\left(\theta_t-\frac{n_t}{z}\right)\right)\right]dt+2\Lambda_tM_t\bar{\sigma}_tX_t^{\alpha}(N\theta_t-X_t)^{\alpha}dB_t^2 
\end{equation}
for $\tau \le t < T$, with explicit solution 
\begin{align}\label{bubblenet}
\beta_t=&\beta_{\tau}e^{-k\int_{\tau}^t \Lambda_sM_s ds}+\int_{\tau}^t \left(-\delta X_s +\lambda N n_s \left(\theta_s-\frac{n_s}{z} \right)\right)\Lambda_sM_se^{-k\int_s^t \Lambda_uM_u du}ds+\notag \\ &+
 \int_{\tau}^t \bar{\sigma}_sX_s^{\alpha}(N\theta_s-X_s)^{\alpha}\Lambda_sM_se^{-k\int_s^t \Lambda_uM_u du}dB_s^2,  \quad \tau \le t < T.
 \end{align}

%\begin{remark}
%Introducing the term $\mathbb{1}_{\left\{\delta \rho_t \le \lambda  n_t \left(1-\frac{n_t}{z}\right)\right\}}$ implies that the bubble does not burst except when $M$ tends to infinity (see Remark \ref{rembubble}). \\ We will need this in the next subsection, where we will show that there exists a flow of equivalent local martingale measures $Q^t$ for $W$ with the property that $W_t^F=\mathbb{E}_{Q^t}[W_T^F|\cal{F}_t]$
%\end{remark}
\begin{remark}
% Notice first that the term $n_t \left(1-\frac{n_t}{z}\right)$ is increasing
%in $n_t$ if and only if $n_t=\sum_k k p_k \rho^k_t<\frac{z}{2}$. This occurs if for example $\rho^k_t<\frac{1}{2}$ for all $k$, which is the case when contagion begins to spread from
%a small fraction of innovators, i.e. from the investors holding the asset at $\tau$. \\
We now consider two different networks, in order to see how the characteristics of the network influence the dynamics of the expected fraction of buyers through $n_t$. In the first one we have a connectivity distribution which is very peaked at the average value $z$ and decaying exponentially fast for $k \gg z$ and $k \ll z$. Examples of this kind of networks are random graph models \cite{ErdosRenyi} and the small-world model of Watts and Strogatz \cite{WattsStrogatz}. In the second one the degree distribution is more right skewed, following for example a power law, as in the Barab\'asi and Albert preferential attachment model \cite{BarabasiAlbert}. From (\ref{Xours1}) and (\ref{nXk}) we can see that the expected contagion between buyers will spread faster in the second kind of network, since the distribution puts more weight on the nodes with higher degree, resulting in a bigger value of $n_t$ in (\ref{nXk}).  \\
As we will notice in the next Section, the more right skewed  is the degree distribution the faster the bubble will build up: this can be seen as an immediate influence of the network on 
the bubble evolution. \\
Looking at (\ref{ourbubble}) there are two opposite forces determining the drift: a negative contribution is given by the speed of decay $-\Lambda_tM_tk\beta_t$, introduced in \cite{JarrowProtter2012}, whereas the term $\lambda N n_t \left(\theta_t-\frac{n_t}{z}\right)- \delta X_t$ is strictly positive when the contagion effects determine the increase of the fraction of buyers. 
%In our setting  the bubble will increase faster then in the Example of \cite{JarrowProtter2012}. \\
When this last term will decay to zero or become negative, the drift will be negative as well: the bubble will revert to zero in expectation. 
\end{remark}
We conclude by showing that there exists a flow $Q^t\in\cal{M}_{loc}(W)$  with Radon-Nykodim derivative process 
\begin{equation}\label{zetanet2}
Z_{t,s}=\frac{dQ^t}{dP}|_{\mathcal{F}_s}=\mathcal{E}\left(\int_0^{\cdot} \alpha_u^{t,1}dB_u^1+\int_0^{\cdot} \alpha_u^{t,2}dB_u^2+\int_0^{\cdot} \alpha_u^{t,3}d\tilde{N}_u\right)_s, \quad s \in [0,T)
\end{equation}
such that 
\begin{equation}\notag
W_t^F=\mathbb{E}_{Q^t}[W_T^F|\cal{F}_t], \quad 0 \le t \le T.
\end{equation}
Taking $\alpha^{t,1}$, $\alpha^{t,2}$ and $\alpha^{t,3}$ in (\ref{alphamu1}), (\ref{alphamu2}) and (\ref{alphamu3}) respectively we only need to show that that $Z$ in (\ref{zetanet2}) is in fact a martingale. 
\begin{proposition}
For each $t \in [0,T]$, $(Z_{t,u})_{u \in [0,T)}$ is a $(P,\cal{F})$-martingale.
\end{proposition}
\textit{Proof}.
We show first that $\mu$ and $\sigma$ in (\ref{munet}) and (\ref{sigmanet}) satisfy Assumption \ref{remM}. \\
Specifically, $$\int_{\tau}^T \mu_s^2ds=\int_{\tau}^T\left(-\delta X_s +\lambda N n_s \left(\theta_s-\frac{n_s}{z}\right)\right)^2ds<\infty$$ by (\ref{inttheta}) and by the fact that
 $$
\left|-\delta X_s +\lambda N  n_s\left(\theta_s-\frac{n_s}{z}\right)\right|< N(\delta\theta_s+\lambda z  \theta_s^2).
$$
Moreover, $$\int_{\tau}^T \sigma_s^2ds=\int_{\tau}^T\bar{\sigma}_s^2X_s^{2\alpha}(N\theta_s-X_s)^{2\alpha}ds\le \left(\int_{\tau}^T\bar{\sigma}_s^4ds\right)^{\frac{1}{2}}\left(\int_{\tau}^TX_s^{4\alpha}(N\theta_s-X_s)^{4\alpha}\right)^{\frac{1}{2}}<\infty$$ by the integrability assumptions (\ref{inttheta}) and (\ref{eqbarsigma}) on $\bar{\sigma}$ and $\theta$, since $0 \le X_s \le N\theta_s$. \\
Finally, by using Feller test it can be seen that the process $X$ does not hit the boundaries $\{0,\theta\}$. Thus
 $\int_{\tau}^T \frac{1}{\sigma_s^4}ds<\infty$ follows by Theorem 2.4 of \cite{MijatovicUrusov2012} and by the integrability assumption on $1/\bar{\sigma}$.
 The thesis follows by Theorem \ref{maintheo}. $\Box$

  \subsection{Numerical simulations}
We now provide a numerical simulation to show how the evolution of the bubble depends on the structure of the network in our model. Specifically, we investigate how the connectivity and the degree heterogeneity of the underlying network influence the dynamics of the bubble. For this purpose, we simulate the dynamics of the bubble evolution specified in the model in (\ref{ourbubble}), 
%\begin{equation}\label{ourbubble2}
%d\beta_t=\Lambda_tM_t\left[-k\beta_t+b \left(-\delta\rho_t + \lambda n_t \left(1-\frac{n_t}{z}\right)\right)\right]dt+\Lambda_tM_tb\bar{\sigma}_t\rho_t(1-\rho_t)dB_t^2,
%\end{equation}
by means of a Monte Carlo method with Euler scheme, taking for simplicity $\Lambda$ and  $\bar{\sigma}$ constant. \\
To include the sudden burst of the bubble we change the dynamics at the moment when the bubble is not growing anymore: when the increase of the bubble stops, the market gets somehow scared, and then a sudden process of pessimistic feeling takes place, leading to the bubble burst that is commonly observed. \\
This has been simulated by increasing the value of $\delta$ and decreasing the value of $\lambda$ in (\ref{ourbubble}) when the bubble remains strictly below its maximum over a certain interval of time.\\
The illiquidity $M$ and the process $\theta$ are supposed to be a geometric Brownian motion, i.e. to satisfy
\begin{align}
&dM_t=M_t(\mu^Mdt+\sigma^MdB_t^3), \quad \tau \le t < T \notag \\
&d\theta_t=\theta_t(\mu^{\theta}dt+\sigma^{\theta}dB_t^3), \quad \tau \le t < T,\notag
\end{align}
where $\mu^{M}, \mu^{\theta} \in \mathbb{R}$ and $\sigma^{M}, \sigma^{\theta} \in \mathbb{R}^+$.\\
We compare two different cases, an Erd\H{o}s-R\'enyi network with Poisson degree distribution 
$$
p_k=\frac{e^{-\tilde{\lambda}}\tilde{\lambda}^k}{k!}, \quad k \in \mathbb{N}, \quad \tilde{\lambda} \in \mathbb{R},
$$ 
and a scale-free network with a power law distribution
\begin{equation}\label{alpha}
p_k \sim k^{-\alpha}, \quad 2<\alpha<3, \quad k \in \mathbb{N}.
\end{equation}
The Erd\H{o}s-R\'enyi network has a degree distribution which is very peaked around the mean degree $z$, whereas the scale-free one, that is well known to better represent real world networks,
 has a much larger right tale, which implies that a bigger number of nodes has a large number of neighbors. \\ 
We take two different values of $\alpha$ in (\ref{alpha}), i.e. $\alpha_1=2.2$ and $\alpha_2=2.5$. obtaining therefore a more connected network (with $z=z_1 \sim 3.2$) and a less connected one (with $ z =z_2 \sim 1.9$). We consider as well two Erd\H{o}s-R\'enyi networks with $z=z_1\sim 3.2$ and $z=z_2 \sim 1.9$, respectively. 
We simulate bubble evolution in these networks, considering the distribution $p_k$ up to a maximum 
degree that corresponds to a network with $50000$ nodes, see paragraph 3.3.2 of \cite{Newman}.\\
For each kind of network we analyze three main quantities:
\begin{itemize}
\item the mean value of the maximum of the bubble;
\item the mean time at which the bubble reaches the maximum (and then it bursts);
\item the value of the bubble at a certain established time, chosen as $t=\tau+0.6$: this is supposed to be an indicator of the speed at which the bubble
developes.                          
 \end{itemize}
 Simulating 10000 trajectories and taking $\delta=0.4$, $\lambda=0.6$, $\Lambda=0.5$, $k=0.1$, $\bar{\sigma}=0.5$, $\tau=0$, $T=3$, $M_0=10$, $\mu^M=0$, $\sigma^M=0.5$, $\theta_0=2$, $\mu^{\theta}=0.2$, $\sigma^{\theta}=0.4$ and $X^k_0=0.02$ for all $k$, we obtain the following results:   \\
 \begin{figure}[h]
 \centering
 \begin{tabular}{|c|c|c|c|c|}
 \hline
 &  scale-free & scale-free & Erd\H{o}s-R\'enyi  & Erd\H{o}s-R\'enyi  \\
 \hline
 mean degree & $3.1987$ & $1.9069$ & $3.1987$ & $1.9069$ \\
 \hline
max  & $2.93\cdot 10^5$ & $2.39 \cdot 10^5$ & $2.73 \cdot 10^5$ & $1.01 \cdot 10^5$  \\
\hline
pos max & $0.73$ & $0.97$ & $1.17$ & $0.78$  \\
\hline
$\beta_{1.6}$ & $1.99 \cdot10^5$ & $9.95 \cdot 10^4$  & $5.36 \cdot 10^4$ & $2.14 \cdot 10^4$ \\
\hline
\end{tabular}
 \caption{Numerical results on bubble evolution in different networks.}
\end{figure}
\vspace{0.5 cm}\\
 One can notice that, as we were expecting, both the mean degree and the degree heterogeneity play a key role in the evolution of the bubble: in particular, both of them are positively correlated with
the steepness of the bubble increase during the ascending phase. \\
It can also be seen that in the Erd\H{o}s-R\'enyi network, i.e. in the less right skewed one, the bubble reaches its maximum later in time: this seems to indicate that the degree heterogeneity leads
to a sooner burst of the bubble. \\
%: in fact, in the case of the more connected networks (the two with $z=3.1987$) even if the bubble grows much more steadly in the scale-free network, in expectation it reaches a slightly higher maximum, later in time, in the Erd\H{o}s-R\'enyi one. \\
The difference in the evolution of the bubble in the two networks can also be seen in Figure 2 and Figure 3, which show five simulated trajectories of the bubble in the free-scale case and in the Erd\H{o}s-R\'enyi case, for $\alpha =2.2$ and $\tilde{\lambda}=3.2$ respectively. Both the networks have the same mean degree $z \sim 3.2$.
In the scale-free network the bubble builds up faster: this is due to the fact that the distribution gives more weight with respect to the Poisson one 
to the nodes with high degree, that are those that in expectation gets faster infected. \\
It is of interest to see the bubble behavior also in the deterministic case, i.e. when $\bar{\sigma}=0$, $\sigma^M=0$ and $\sigma^{\theta}=0$, see Figure 4. In the more connected and right skewed network the bubble builds up faster, and bursts faster as well. The greatest maximum is reached in the Erd\H{o}s-R\'enyi network with $\tilde{\lambda}=3.2$, where the bubble builds up slower than in the scale-free network but it reaches a bigger value in average. As before, we see that also the connectivity
leads to a faster growth and to a sooner burst. 
\begin{figure}
\centering
 \includegraphics[scale=0.6]{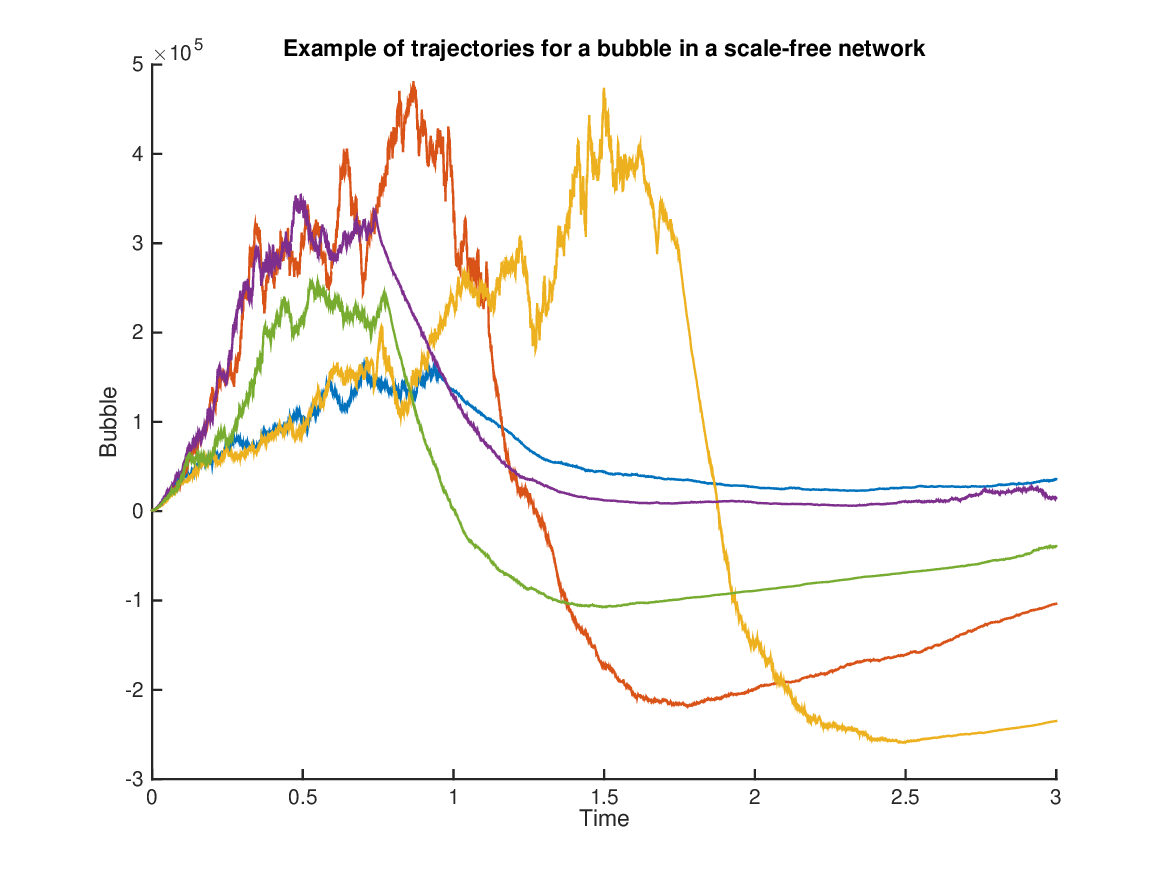}
 \caption{Scale-free network.}
\end{figure}
\begin{figure}
\centering
 \includegraphics[scale=0.6]{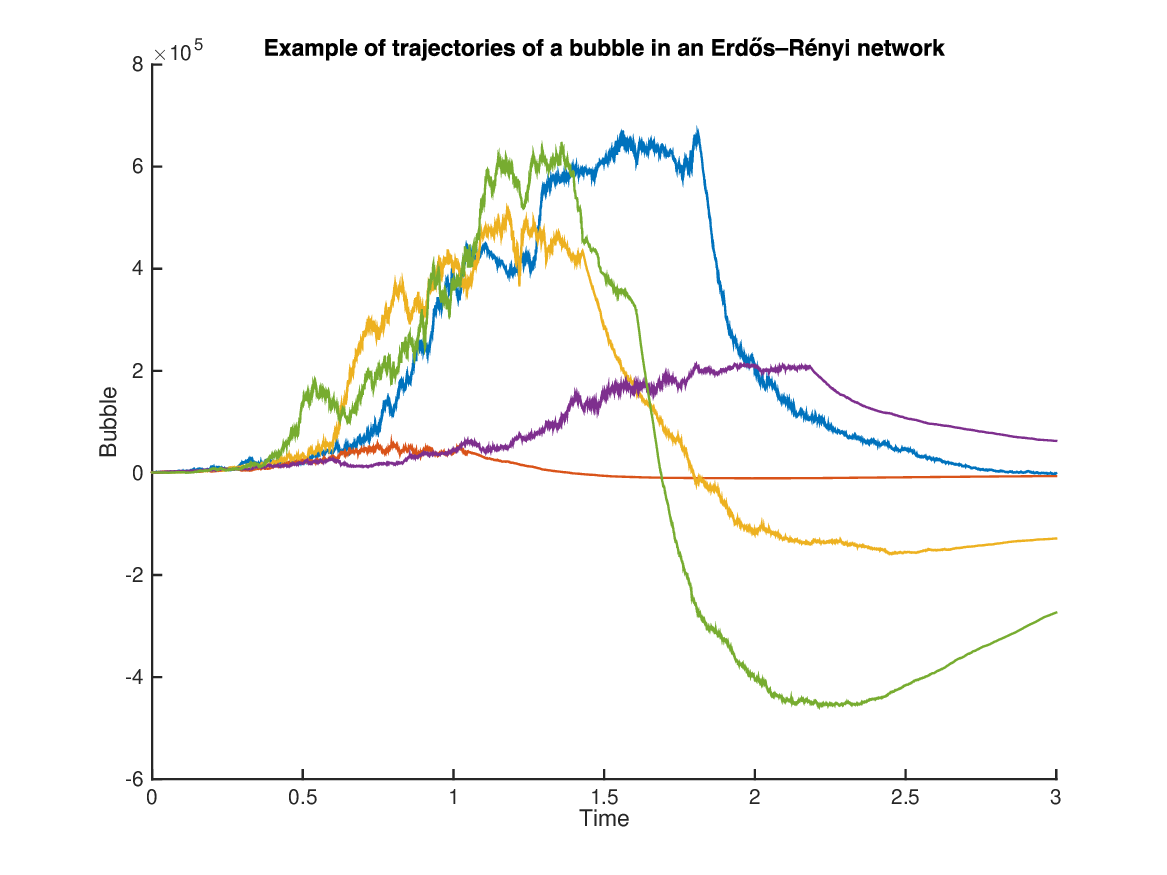}
 \caption{Erd\H{o}s-R\'enyi  network.}
\end{figure} 
\begin{figure}
 \centering
 \includegraphics[scale=0.6]{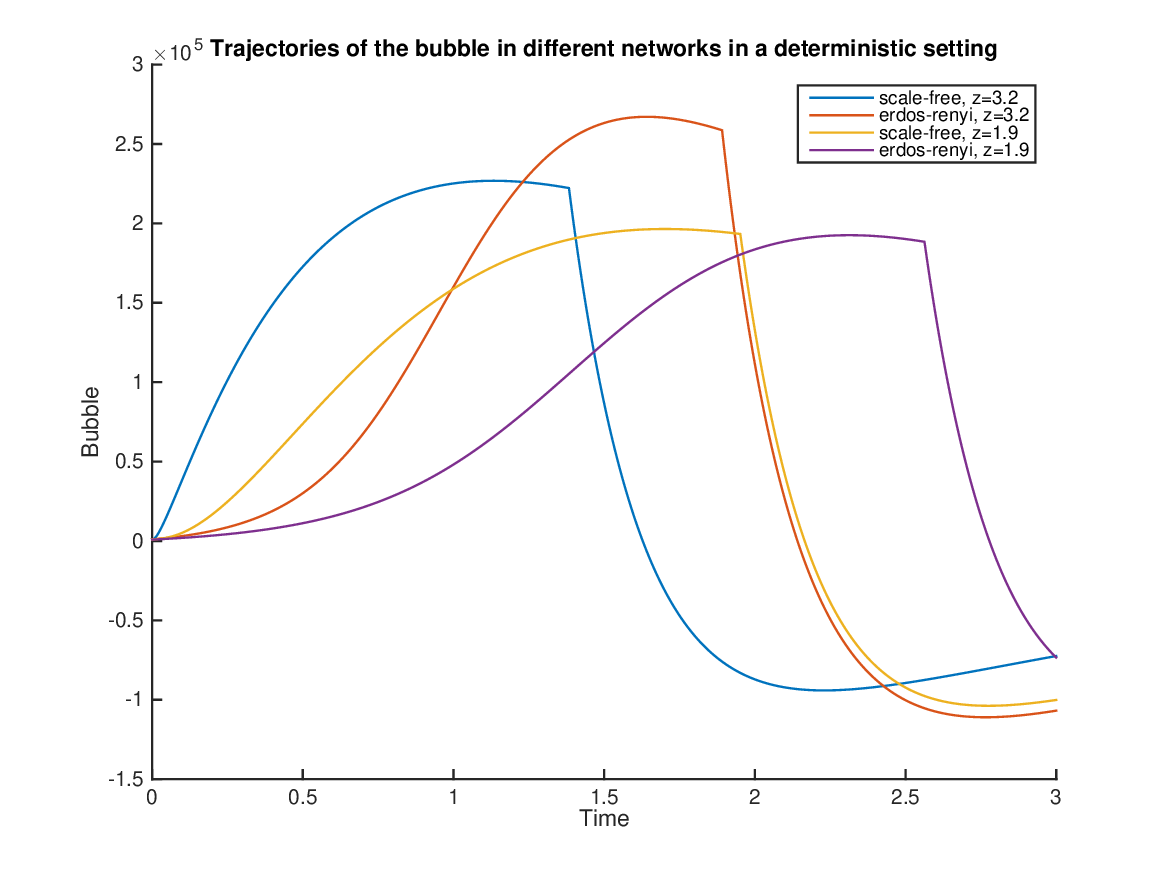}
 \caption{Deterministic example.}
\end{figure}  
% To finish, here is the Matlab function that simulates the trajectories of the bubble.
%\lstinputlisting{montecarlobubble.m}
 
 \newpage

\bibliography{bib}
\bibliographystyle{plainnat}
\end{document}